# The Effects of Access to Credit on Productivity: Separating Technological Changes from Changes in Technical Efficiency

Nusrat Abedin Jimi[1], Plamen Nikolov[1,2,3],
Mohammad Abdul Malek[4], and Subal Kumbhakar[1]

Improving productivity among farm enterprises is important, especially in low-income countries where market imperfections are pervasive and resources are scarce. Relaxing credit constraints can increase the productivity of farmers. Using a field experiment involving microenterprises in Bangladesh, we estimate the impact of access to credit on the overall productivity of rice farmers, and disentangle the total effect into technological change (frontier shift) and technical efficiency changes. We find that relative to the baseline rice output per decimal, access to credit results in, on average, approximately a 14 percent increase in yield, holding all other inputs constant. After decomposing the total effect into the frontier shift and efficiency improvement, we find that, on average, around 11 percent of the increase in output comes from changes in technology, or frontier shift, while the remaining 3 percent is attributed to improvements in technical efficiency. The efficiency gain is higher for modern hybrid rice varieties, and almost zero for traditional rice varieties. Within the treatment group, the effect is greater among pure tenant and mixed-tenant farm households compared with farmers that only cultivate their own land.

**JEL Classification** E22, H81, Q12, D2, O12, O16

---

[1] Department of Economics, State University of New York, Binghamton
[2] Harvard University Institute for Quantitative Social Science
[3] IZA Institute of Labor Economics
[4] Research and Evaluation Division, BRAC and Kyoto University, Japan

# 1.  Introduction

Subsistence farms in developing countries face a difficult environment characterized by a high degree of risk, credit constraints, a lack of financial markets, high input costs, and time-inconsistent preferences (Duflo 2006). These factors shape smallholder farming practices and performance, as well as production and investment decisions (Stiglitz, Emran and Morshed 2006; Bruhn, Karlan and Schoar 2010; Karlan et al. 2014). Provision of agricultural credit at a subsidized interest rate mulla be an effective tool for enhancing the production of rural farms. Relaxing the credit constraint for farming enterprises could lead to greater adoption of modern inputs and improved ability to turn inputs into outputs, both of which boost productivity. Productivity and efficiency underscore the organizational capacity of subsistence farmers to deal with external shocks, and have far-reaching implications in terms of ensuring their sustainable livelihood (World Bank 2004). Therefore, understanding the relationship between credit constraints and farm productivity and efficiency has crucial policy implications. If the relaxation of credit constraints produces efficiency improvements, policymakers need to account for these additional benefits of credit programs.

In this study, we examine how access to credit influences farmer productivity, and whether the effects on output come from changes in technology and/or from increased efficiency.[1] We do so by using survey data from a field experiment[2] that exploits the random assignment of credit services to agricultural enterprises by the Bangladesh Rural Advancement Committee (BRAC)[3] and employing a stochastic production frontier model.[4] First, we examine the impact of credit access on the productivity of rice-producing farms.[5] Then, we disentangle the overall productivity effect into technological change and changes in efficiency. In addition to identifying the impacts of credit access on productivity, technological change, and technical efficiency, we examine how the change in efficiency varies in relation to several demographic

---

[1] We define productivity as yield per unit of land (kilogram of rice per decimal of land). A decimal (also spelled decimel) is a unit of area in India and Bangladesh approximately equal to 1/100 acre (40.46 m²); 247 decimal=1 hectare.
[2] As per the taxonomy presented by Harrison and List (2004).
[3] The largest NGO in Bangladesh.
[4] The conventional production function approach does not allow us to separate technological change (frontier shift) and efficiency improvements from the overall productivity effect. We use the stochastic frontier model because it allows us to decompose these two effects, we use this approach as a tool to answer our research question.
[5] Hossain et al. (2018), Hossain et al. (2016), and Malek et al. (2015) examine the impact of the BCUP program on asset holdings, aggregate welfare, and wage employment.

and farm characteristics.

Previous empirical studies have found that provision of microcredit to farmers can boost their yield and productivity (McKernan 2002; Chirkos 2014; Hussain and Thapa 2012; Rahman, Hussain and Taqi 2014), as well as leading to higher income-generating activities in developing countries (Karlan and Zinman 2009; Kondo et al. 2008; Montgomery and Weiss 2011).[6,7] For example, using data from a survey carried out in rural Bangladeshi villages, McKernan (2002) estimated the impact of household participation in three microcredit programs (BRAC, BRDB's RD-12 program, and Grameen Bank). Measuring the total and noncredit effects of microcredit program participation on productivity, McKernan (2002) found a significant positive effect of participation in the credit program on productivity among self-employed enterprises.

In addition to the empirical literature examining the link between credit expansion and productivity, other studies have examined whether credit influences farm efficiency. Conceptually, relaxing credit constraints has an ambiguous effect on the technical efficiency of farming enterprises. On the one hand, less credit-constrained farms can procure more inputs and more easily cover operating expenses in the short run (Singh, Squire, and Strauss 1986; Blanchard et al. 2006), enabling them to make better investments in the long run (Hadley et al. 2001; Blanchard et al. 2006; Davidova and Latruffe 2007; Guirkinger and Boucher 2008; Karlan et al. 2014).[8] Credit can also mitigate consumption risk and enable greater adoption of modern inputs by small, risk-averse farms (Liu and Zhuang 2000; Easwaran and Kotwal 1989). On the other hand, if farmers face other constraints or market imperfections (Jack 2013), such as lack of access to insurance, lack of markets, or high input costs, then access to credit will not translate in to higher productivity and technical efficiency.

Recent empirical studies have attempted to measure inefficiency in agricultural enterprises and examine the factors underlying this inefficiency (Anang, Backman and Sipiläinen 2016; Islam, Sumelius and Bäckman 2012; Bravo-Ureta et al. 2007). However, the findings of these studies are largely based on observational designs, and the determinant factors are not based on any exogenous changes. Furthermore, many studies have produced conflicting results. For instance, using parametric efficiency analysis, Taylor et al. (1986) and Brummer (2000) found a negative relationship between relaxed credit constraints and the efficiency of farmers. However, other studies focused on the Philippines, West Bengal, Pakistan, and Bangladesh found relaxed credit constraints to be an important and beneficial determinant of farm efficiency (Martey, Wiredu and Etwire 2015; Islam, Sipilainen and Sumelius 2011).

---

[6] Numerous other experimental field studies, documented in Banerjee, Karlan and Zinman (2015), Banerjee (2013), and Roodman (2014), have examined how the availability of microcredit affects other important outcomes, such as business size and profits (Banerjee, Karlan and Zinman 2015), income composition (Banerjee, Karlan and Zinman 2015), stock of household durables (Attanasio et al. 2015), occupational choice, business scale, and risk management (Banerjee, Karlan and Zinman 2015), female decision-making power (Angelucci, Karlan and Zinman 2015), and happiness and trust (Angelucci, Karlan and Zinman 2015).

[7] De Janvry, Sadoulet, and Suri (2017) reviewed all of the recent experimental field studies on agricultural inputs in developing countries. We focus on credit as an input and its impact on productivity and technical efficiency. The impact of other production inputs on productivity has also been examined in relation to inputs such as credit (de Mel, McKenzie, and Woodruff 2008, McKenzie and Woodruff 2008), capital (Karlan, Knight and Udry 2015), labor (Shearer 2004), information (Beaman and Magruder 2012), monitoring (Nagin et al. 2002), and managerial practices (Bloom and Van Reenen 2010, Karlan and Valdivia 2011; Drexler, Fischer, and Schoar 2014).

[8] Credit has been shown to affect the risk-taking behavior of producers (Boucher, Carter and Guirkinger 2008; Eswaran and Kotwal 1990), thereby affecting technology choices and adoption by farmers. The timing of the investment decision can also play an important role in one's risk preferences (Nikolov 2018).

Building on existing studies that have examined how credit influences productivity, we investigate how credit expansion, as a result of a subsidized interest rate,[9] influences productivity gains via two distinct channels: technical change (frontier shift) and change in technical efficiency. Using a stochastic frontier approach, we separate the frontier shift effect from the efficiency effect.

We find that relaxing the credit constraint has a significant positive impact on rice production, both in relation to frontier shift and technical efficiency. We find a positive impact from access to credit on total rice output, specifically high-yielding variety (HYV) rice and hybrid rice, but no impact on traditional rice varieties. We find that, relative to the baseline, credit access increases overall productivity by, on average, approximately 14 percent, with the greatest impact on modern hybrid rice growing farms. After decomposing the overall output effect into frontier shift and efficiency change effects, we find that around 11 percent of the overall productivity gain comes from technological change, or frontier shift. In terms of technical efficiency, small-scale farms with access to subsidized credit are, on average, 3 percent more efficient than farms[10] without credit access (which, relative to the average baseline rice yield of 18 kilograms per decimal, implies approximately half a kilogram less lost output as a result of inefficiency). This positive effect is even more pronounced among producers of hybrid rice varieties, who exhibit an efficiency gain of, on average, 9 percent. Moreover, we find different impacts among marginal and tenant farm households.[11] Our results show that among the farms with credit access, enterprises with less than 50 decimals under cultivation are, on average, 3 percent less efficient than larger farms. We also find strong evidence of a positive effect of credit access (at the 95th percentile level) on efficiency for tenant farm households compared with pure owner farms.

---
[9] Although we rely on an exogenous change in the price of borrowing as a result of the fact that the treatment group obtains access at a subsidized rate, other studies have examined exogenous changes in other aspects of microcredit programs such as microcredit access (Banerjee et al. 2015; Crepon et al. 2015), loan maturity (Karlan and Zinman 2008), and loan eligibility (Karlan and Zinman 2009).
[10] In this study, we use the term "farm households" interchangeably with "poor rice farmers".
[11] Tenant farm households are farms that cultivate other people's land, either through sharecropping or renting, or both.

Our findings highlight the likely mechanisms explaining the positive impacts of microcredit access on productivity and efficiency. In the absence of insurance and credit markets, credit-constrained households are more likely to continue their conventional farming practices. Enhanced access to credit enables farm households to adopt more productive crop varieties and utilize complementary production inputs in a more timely manner. Credit can also boost farms' potential to manage and allocate their resources more effectively, which also results in increased output. We find that the adoption of modern hybrid rice varieties is significantly higher, on average, among households that have credit access. Furthermore, households with access to credit procure significantly more pesticides, which are essential in ensuring stable yields of hybrid rice varieties.[12] We find larger productivity gains among producers of modern rice varieties and almost no gains among producers of traditional rice varieties. One explanation for this difference might be that modern rice varieties offer greater potential yields, but also require more complementary inputs, and the timely application of those inputs, which farmers find easier to manage when they have access to credit. Although our study is limited to the impact of credit rather than the combined impact of credit and extension services,[13] our analysis shows that farmers with access to credit are more likely to be familiar with and able to discuss crop choices, input choices, and farm practices with agricultural extension service officers and providers than those without access (see Table A5 in the Appendix).[14]

This study makes three important contributions to the empirical literature. First, relative to previous empirical studies, we identify more credible, causal impacts of credit access on the productivity and efficiency of small farm enterprises in a low-income country context by exploiting the experimental design of the BCUP program, augmenting our analysis with a stochastic frontier approach. Specifically, the random assignment of microcredit access ensures that the technologies of the two groups (treatment and control), which we use in the stochastic frontier analysis, remain fixed at the baseline. Second, our study complements previous field experiments (Banerjee et al. 2015). Although previous studies have examined the impact on productivity (McKernan 2002; Chirkos 2014; Hussain and Thapa 2012; Rahman, Hussain and Taqi 2014), we contribute to the empirical literature by examining the specific source of the overall productivity gain, that is, whether it comes from a frontier shift or from improved efficiency. Third, recent economic studies have found that various forms of scarcity can influence optimizing behavior among the very poor (Shah, Mullainathan and Shafir 2012; Shah, Shafir and Mullanathan 2015). Adding to this strand of the literature, we examine how credit can influence efficiency among the poor. Because the poor already operate and make decisions under conditions of significant scarcity of resources, shedding light on how to improve the efficiency of their farming enterprises has important welfare implications.

---

[12] The timely and repeated use of pesticides is very important in ensuring higher returns from modern hybrid rice varieties.
[13] The BCUP program included complementary extension services in the initial years. However, BRAC ceased to provide extension services in 2012 because of high attrition rates and high recovery costs (Hossain et al. 2018).
[14] For simplicity, we have not modeled risk in this study.

The rest of the paper is organized as follows. Section 2 presents the program design, data sources, and summary statistics. Section 3 describes the conceptual framework and the channels through which credit influences the two study outcomes. The empirical strategy is described in Section 4. Section 5 presents the main results, and Section 6 concludes with a discussion of the findings.

## 2. Project Background

### 2.1 The BCUP Credit Program

In 2009, BRAC introduced a Tenant Farmers Development Project known as Borga Chashi Unnayan Prakalpa (BCUP). The project was initiated with Tk. 5,000 million (USD 70 million) as a revolving loan from Bangladesh Bank, the central bank of Bangladesh, at a monthly interest rate of 5 percent, the rate at which commercial banks can borrow funds from the central bank. Funding was initially offered for three years, with the aim of providing credit to 300,000 farmers.

The main objective of the BCUP program was to reduce the dependence of tenant farmers on high-cost informal markets for financing their working capital needs. Tenant farmers are typically bypassed by conventional microfinance institutions and the formal banking sector, resulting in a lack of working capital, and thus restricted access to inputs and lower productivity (Hossain and Bayes 2009). By reducing the credit constraints faced by these farmers, the BCUP program aimed to significantly improve farm productivity, and thus the livelihoods of rural small-scale farm households in Bangladesh.

BCUP provides a customized credit service based on the proprietary composition of the recipient farms, that is, pure tenant, mixed tenant, or pure owner. Loans are provided at a reduced fixed interest rate of 10 percent per year (see Figure 3). If a farmer cannot repay an installment by the due date, he/she must pay additional interest with the remaining installments. The effective rate of interest is 19 percent on a declining balance basis, which is still lower than the 27 percent charged by other microfinancing programs in Bangladesh.[15] The loan amount ranges from a minimum of 63 USD to a maximum of 1,500 USD (Tk. 5,000–120,000), the duration is 6–10 months, the grace period is one month, and repayment is by monthly installment. BCUP targeted all 484 upazilas (sub-districts) of Bangladesh in successive phases. According to BRAC Microfinance administrative data, the BCUP program disbursed 8 billion USD in loans to about 700,000 farmers between its launch in 2009 and June 2018.

Households are selected for loan disbursement based on several stages of verification. The first stage entails the initial selection of members. Members are selected by assessing each household against the BCUP eligibility criteria and familiarizing farmers with the BCUP program and its terms and conditions.[16] In the second stage, a farmer is assigned to the nearest village organization (VO) given that he/she agrees to the terms and conditions of the BCUP. Stage three entails the collection of more detailed information about members. In the fourth and final stage, the list of members is finalized after verification by a branch manager, who determines the eligibility of the members who were initially selected.

After this selection process, new members are formally admitted and attend an orientation meeting. An important feature of the BCUP program is the formation of the village organization (VO) and its use as a platform for service delivery. A total of four to eight five-member teams, that is, 20 to 40 farmers, consists a VO. The VO members meets once a month at a set time on a fixed day, and the BCUP program organizer attends the VO meeting to discuss loan proposals and collect repayment installments, dues, and savings deposits.

---

[15] As per the rules of the Microcredit Regulatory Authority (MRA) of Bangladesh Bank, NGOs can charge up to a maximum of 27 per cent interest on declining balances through their microfinance operations.
[16] The eligibility criteria for the BCUP program were: 1) The farmer has a National ID card; 2) The age of the farmer is between 18 and 60 years; 3) The education level of the farmer is no higher than SSC; 4) The farmer must have been a permanent resident of the area for at least three years; 5) The farmer has at least three years of prior experience in farming; 6) The land holding must be between 33 decimals and 200 decimals; 7) The farmer cannot be an MFI (Micro Finance Institution) member; and 8) The farmer must be willing to accept credit from BCUP.

The BCUP program included complementary extension services in the initial years when BRAC's agricultural development officers attended the monthly VO meetings to provide information and advice on modern cultivation systems and farm management. However, because of high attrition rates and high recovery costs, BRAC ceased to provide extension services in 2012. Therefore, this study is limited to the impact of credit access, rather than the combined impact of credit access and extension services. Table 1 presents the summary statistics of the baseline household composition of program participants in relation to the various inputs used and output in the form of rice production.

[Table 1 about here]

## 2.2 Experimental Design and Baseline Survey

The BCUP program was established under a clustered randomized control trial design. Initially, the program identified 40 potential sub-district/branch[17] offices for program scale-up in 2012. The research team randomly selected 20 treatment branch offices for intervention, while the other 20 branches were designated as control branch offices. Then, we randomly selected six of the 10–12 villages within an eight-kilometer radius of each BCUP branch office. The eight-kilometer radius was chosen because BRAC branch offices usually operate within this area for administrative purposes. The sub-district/branch is the first unit of randomization, followed by the village/community. As each branch is located in a different sub-district, and each sub-district is a separate government administrative unit with a well-known geographical boundary, contamination between the treatment and control BCUP branches is unlikely. Figure 2 provides a spatial overview of the treatment and control areas. It can be seen that most of the treatment branches were sufficiently distant from control branches.[18]

We conducted a household-level census in all 240 villages to identify eligible households. The census covered a total of 61,322 households, of which 7,563 households fulfilled the program eligibility criteria[19] and were willing to accept agricultural credit.[20] Then, we randomly selected 4,301 of these households for detailed data collection, 2,155 households from treatment villages and 2,146 households from control villages.[21] The baseline survey on various inputs and rice output was conducted in 2012,[22] and a short-term follow-up survey was carried out in 2014. Figure 1 shows the experimental study design. Households in the treatment units were provided with access to credit of up to 120,000 Tk. (≈1500 USD). Figure 3 shows the features of the program.

---

[17] The sub-district (upzila) is an administrative unit in Bangladesh. There are 491 sub-districts in Bangladesh.
[18] A few branches in the southern region were exceptions. For the southern region branches, GIS mapping (see Figure A1 in the Appendix) was undertaken and the results were forwarded to the program administrators so that they could continue to expand the number of treatment intervention branches within the appropriate areas while avoiding incursions into the control areas. Because the BCUP program administrators were aware of the status of each branch in the study, it was unlikely that the program officers would disburse loans in a control branch (Malek et al 2015).

[Figure 1 about here] [Figure 2 about here] [Figure 3 about here]

With random assignment of study subjects to one of the two groups, the baseline census characteristics should be, on average, the same across the treatment and control groups, apart from sampling variations. Columns 1 and 2 in Table 2 show the baseline means of the variables for the control and treatment groups, respectively.[23] We tested the equality of the means by random assignment of credit access, and column 3 in Table 2 presents the associated $p$-values. We found that almost all of the 26 differences between the control and treatment groups had a $p$-value of less than 0.10, except for female-headed households, which suggests that the baseline mean characteristics of the two groups are statistically similar.[24] We also performed a joint test of orthogonality to test for baseline balance. The result of the joint significance test is shown in the final row of Table 2. These findings are consistent with the successful implementation of random assignment of subjects.

[Table 2 about here]

We also checked whether households in the treatment group dropped out of the study at a different rate to those in the control group (see Table 3). A substantial difference in attrition rates could result in biased study results if it is related to the initial assignment of subjects. We found an attrition rate of around 10 percent in the panel data used in the field experiment, and no significant difference between the attrition rates in the treatment group (11 percent) and control group (9 percent) for rice-producing farm households (see column 1).[25]

---

[19] Described in Section 2.1.
[20] Willingness to accept credit is measured by a 'Yes' or 'No' answer in response to the question of whether a respondent is inclined to accept credit from the BCUP program.
[21] We adopted a simple random sampling method to select households from each village. The survey covered 4,301 households, of which 2,155 were in treatment areas and 2,146 were in control areas.
[22] Following the baseline survey, we forwarded the list of treatment branches to the BRAC-BCUP administrators, whereupon BRAC launched the BCUP program in the treatment branches. The program organizers visited all the villages to locate potential borrowers based on the eligibility criteria.
[23] For balancing checks, we restricted our sample to the rice producing farm households surveyed in 2012 (3,292 households).
[24] The results of the balancing tests by rice variety are presented in Appendix A (Tables A1 and A2).

[Table 3 about here]

## 2.3  Data

We used baseline data and follow-up survey data at the household level from the BCUP program. A total of 4,301 households (2,155 in treatment areas and 2,146 in control areas) were randomly selected to participate in a quantitative baseline survey in 2012, and a follow-up survey was conducted in 2014 (Hossain et al. 2018; Malek et al. 2015). For simplicity, we focused on rice-producing farms.[26] The data include economic and demographic variables relating to farm households, as well as inputs and output in terms of rice production. Our input variables included land (decimals), labor (days), ploughing land in preparation for planting (number of times), seed (kilograms), irrigation (hours), fertilizer (kilograms), and pesticide (number of times used).

## 3.  Conceptual Framework: Credit Use, Change in Technology and Efficiency

Once a farm enterprise obtains access to credit ($Z$), its output can be affected through various channels. In the following sections, we denote access to credit by a farm household by $Z_i$, which takes a value of 1 if the farm household is assigned to the treatment group (eligible for credit under the BCUP program) and 0 otherwise. Being in the treatment group can increase the use of inputs by a credit-constrained farm, and can also lead to a shift in the production frontier. Meanwhile, it can also increase output by improving efficiency. There might also be a synergistic effect involving both technological change and efficiency improvement. We represent the general production function as:

---

[25] In column 2 of Table 3, we present results of the regression of attrition in the follow-up survey on treatment dummy and household covariates. and find no evidence that treatment assignment is statistically significantly related to household attrition status.
[26] Rice is a major crop produced in Bangladesh, and almost all of the farm families in the country grow rice. Rice is cultivated on 75 percent of the country's cropland (Ganesh-Kumar, Prasad and Pullabhotla 2012), and is the primary source of income and employment for nearly 15 million farm households in Bangladesh (Bangladesh Bureau of Statistics 2008).

$$lnY = lnf(X,Z) - u(Z) + v, \qquad (1)$$

where $lnY$ is the log of rice output, $X$ is the vector of inputs including land, labor, machinery, seed, irrigation, fertilizer, and pesticides. $v$ is noise, and $u$ is technical inefficiency. Our primary objective is to examine the effect of Z on output while leaving the input vector unchanged, and decomposing the effect into a frontier shift and a change in efficiency.

## 4. Estimation Strategy

We estimate the impact of access to credit on productivity and efficiency by comparing the average outcomes of the treatment and control groups. Therefore, our estimates are based on the initial treatment assignment irrespective of households' actual enrollment or participation in the BCUP program. We start by estimating the impact of credit expansion to farm households on the use of production inputs and adoption of modern rice varieties. Then, we examine the impact of credit expansion on productivity. A Cobb–Douglas (CD) production function is used to represent the production technology. To estimate the function, we initially use the ordinary least squares (OLS) method, and then discuss the problem this approach presents in relation to decomposing the total effect on output into technological change and efficiency improvement effects. Next, the stochastic production frontier approach is presented, and we explain how we use this model as a tool to disentangle the two effects, frontier shift and efficiency improvement.[27]

### 4.1 Effect of Credit Access on Input Use and Adoption of Modern Rice Varieties

Before examining the overall impact of credit access on productivity, we check whether credit belongs to the input set by examining the impact of credit access on the use of different inputs. As mentioned earlier, $T_i$ is the treatment indicator. The difference in outcomes between the treatment and control groups (i.e., households with credit access and those without) is known as the intent to treat (ITT) effect, and is captured by the following OLS regression:

$$Q_i = \delta_0 + \delta_1 T_i + \vartheta_i, \qquad (2)$$

where $Q_i$ is the outcome variable (use of land, labor, fertilizer, and pesticides and adoption of modern hybrid rice varieties), $T_i$ is an indicator of assignment to either the treatment or control group, and $\vartheta_i$ is the error term. The parameter of interest is $\delta_1$, which captures the ITT effect — the average effect of simply being offered access to the credit program—on changes in the outcome variables twenty-four months after the start of the intervention. We cluster the standard errors at the branch level to account for intra-cluster correlation.

---

[27] It is important to note that our analysis only covers a partial equilibrium effect and does not capture first-order general equilibrium effects. Moreover, the coverage of the BCUP credit program is not large enough to create a village-level effect. As noted in Section 2.1, the BCUP program uses the VO as the platform for service delivery. Members are grouped into teams of five, and three to eight teams consisting of 15 to 40 members form a village-level tenant farmer association. BCUP program administrative data from 2012 suggest that sometimes the number of participants in a village is insufficient to form an association, and so two or three villages must be combined. Therefore, although the theoretical maximum number of BCUP participants from a village can be as many as 40, in reality the number is much lower, and is not a large proportion of the total number of farm households in a village.

## 4.2  Effect of Credit Access on Productivity: Overall Effect

To formalize our analysis, we use the indicator variable $Z_i$ to represent credit access and rewrite equation (1) (using the CD production function) as:

$$lnY_i = \beta_0 + \sum_j \beta_j \, lnX_{ji} + \gamma_1 Z_i - u_i(Z_i) + v_i, \qquad (3)$$

where $i$ denotes each rice producing farm household, $lnY_i$ is rice output per decimal (in log form), $\ln X_{ji}$ is the log of input variable $j$ per decimal of farm $i$, $v_i$ is noise, and $u_i(Z_i)$ is the inefficiency term. $X_{ji}$ includes land (decimals), labor (days), ploughing land in preparation for planting (number of times), seed (kilograms), irrigation (hours), fertilizer (kilograms), and pesticide (number of times).[28]

To explore the consequences of applying OLS in the presence of inefficiency, we further rewrite the equation as:

$$lnY_i = \beta_0 + \sum_j \beta_j lnX_{ji} + [\gamma_1 Z_i - E(u_i(Z_i))] + v_i$$

---

[28]Note that coefficient of $lnL$ in specification (3) can be expressed as $RTS - 1$. A positive coefficient indicates RTS of land is greater than 1 and a negative coefficient means RTS of land is less than 1.

$$\equiv \beta_0 + \sum_j \beta_j \ln X_{ji} + \gamma Z_i + \epsilon_i, \qquad (3a)$$

where $\epsilon_i = \{v_i - [u_i(Z_i) - E(u_i(Z_i))]\}$, and $\gamma Z_i = \gamma_1 Z_i - E(u_i(Z_i))$. By construction, $\varepsilon_i$ has a mean of zero, and so OLS can be used to estimate equation (3a).

Therefore, as shown in equation (3a), the $Z_i$ term has two effects, represented by $[\gamma_1 Z_i - E(u_i(Z_i))]$. The first term is the direct effect on technology, while the second term captures the effect on efficiency. If inefficiency is not explicitly modeled, the coefficient of $Z_i$ in equation (3a) will capture the mean overall effect of expanded credit access on productivity.[29] In other words, if inefficiency is not explicitly included and $E(u_i(Z_i))$ is approximately linear in $Z_i$ (that is, $E(u_i(Z_i)) = \gamma_2 Z_i$ so that $\gamma = \gamma_1 + \gamma_2$), the coefficient of $Z_i$ ($\gamma$) will capture both the technology change (frontier shift, $\gamma_1$) and the change in efficiency ($\gamma_2$). The estimated coefficient of $Z_i$ in equation (3a) does not enable us to disentangle the frontier shift and efficiency improvement effects.

### 4.3 Effect of Credit Access on Productivity: Separating the Frontier Shift Effect from the Efficiency Effect

In this subsection, we use the stochastic frontier approach instead of the distribution-free[30] approach used in Subsection 4.2 to separate the frontier shift effect from the inefficiency effect.

We specify our production model as follows:
$$\ln Y_i = \ln Y_i^* - u_i(Z_i), \qquad u_i(Z_i) \geq 0 \qquad (4)$$
$$\ln Y_i^* = \ln f(X_i; \beta) + v_i. \qquad (5)$$

Equation (5) defines the stochastic production frontier function. Note that the error is composed of two terms – the inefficiency term $u_i(Z_i)$ and the noise term $v_i$. For a given level of $X$, the frontier gives the maximum level of output ($Y_i^*$), and is stochastic because of the presence of $v_i$. Rearrangement of equation (4) gives $\exp(-u_i(Z_i)) = \frac{Y_i}{Y_i^*}$ (the ratio of actual output to maximum possible output), and the value of $(1 - \exp(-u_i(Z_i))) \times 100$ is the percentage by which actual output falls short of the maximum possible output. Since $\exp(-u_i(Z_i)) \approx 1 - u_i(Z_i)$, $u_i(Z_i)$ is referred to as the technical inefficiency of farm household $i$. The presence of inefficiency gives rise to a composite error term $[v_i - u_i(Z_i)]$, which is negatively skewed because $u_i(Z_i)$ is one-sided.[31] We perform a simple OLS residual test to check for skewness of the error term, and thus the appropriateness of using the stochastic frontier specification. We also run a sample moment-based test following Coelli (1995). Both results reject the null hypothesis of no skewness in the OLS residuals in the baseline, suggesting the presence of inefficiency.

---

[29] One might argue that the effect of credit access on the production frontier operates through inputs: credit enables poor farmers to use pesticides and fertilizer, and buy modern seed varieties in a timely manner, thereby affecting the production frontier. However, the relationship might be linear for some inputs and nonlinear for others. For simplicity, we are trying to find the overall effect of credit access. Therefore, we add credit access as a separate factor in the production frontier (that is, $\gamma_1 Z_i$) rather than examining the effect of credit through inputs.

[30] In this approach, the estimation results do not impose any distributional assumption on $u_i(Z_i)$. However, the major drawback of this approach is that the inefficiency effect cannot be separated from the noise ($Z_i$) if the inefficiency is i.i.d. (a function of $Z_i$).

As before, we use a simple CD technology function to represent $f(.)$. Additionally, we assume that the inefficiency term $(u_i(Z_i))$ follows a half-normal distribution. We parameterize $u_i(Z_i)$ as a function of the treatment assignment variable $(Z_i)$, and therefore allow the randomly assigned access to credit $(Z_i)$ to affect the expected value of the inefficiency. We then apply the maximum likelihood method to estimate the model parameters (parameters in $f(.)$) and inefficiency in the single-equation approach, following Kumbhakar, Wang, and Horncastle (2015). Specifically, our model is:

$$lnY_i = lnY_i^* - u_i(Z_i), \quad u_i(Z_i) \geq 0 \quad (6)$$
$$lnY_i^* = ln\,X_i\beta + \gamma_1 Z_i + v_i, \quad (7)$$
$$u_i(Z_i) \sim N^+(0, \sigma_u^2(Z_i)), \quad (8)$$
$$\sigma_u^2(Z_i) = \exp(\delta_0 + \delta_1 Z_i), \quad (9)$$
$$v_i \sim i.i.d\ N(0, \sigma_v^2), \quad (10)$$

where $X$ is the vector of inputs[32] and $\beta$, $\gamma_1$, $\delta_0, \delta_1$, and $\sigma_v^2$ are the parameters to be estimated. $\gamma_1$ captures the impact of credit access on the frontier shift (technological change), while $\delta_1$ represents the effect of credit access (rather than the marginal effect of credit) on inefficiency. After estimating the model parameters and the (in)efficiency index under the single-equation approach, we obtain the marginal impact of credit access $(Z_i)$ on the expected value of the inefficiency $u_i(Z_i)$ from $E(u_i(Z_i = 1)) - E(u_i(Z_i = 0))$, where $E(u_i(Z_i)) = \sqrt{2/\pi}\, \sigma_u(Z_i) = \sqrt{2/\pi}\, (.5(\exp(\delta_0 + \delta_1 Z_i)))$. Therefore, the marginal effect of $Z_i$ is decomposed into the frontier shift effect (given by the coefficient of $Z_i$ in the production frontier, $\gamma_1$) and the technical efficiency effect obtained from $E(u_i(Z_i = 1)) - E(u_i(Z_i = 0))$. The sum of these two values gives us the overall effect of $Z_i$ on output, holding all other inputs unchanged. It is important to mention that the sum of the two effects does not necessarily equal $\gamma$ in (3a) unless $E(u_i(Z_i))$ is approximately linear.[33] Also, note that although we model the credit access ($Z_i$) as a determinant of inefficiency, we do not present any analysis on the variance of the noise term in this paper for simplicity.[34]

---

[31] For a production-type stochastic frontier model with the composite error $v_i - u_i(Z_i), u_i(Z_i) \geq 0$ and $v_i$ distributed symmetrically around zero, the residuals from the corresponding OLS estimation should skew to the left (that is, negative skewness) regardless of the distribution function of $u_i(Z_i)$ in the model estimation after pretesting. Thus, a test of the null hypothesis of no skewness can be constructed using the OLS residuals. If the estimated skewness has the expected sign, the rejection of the null hypothesis provides support for the existence of one-sided error.
[32] Inputs are in log form and include land (decimals), labor (days), ploughing land in preparation for planting (number of times), seed (kilograms), irrigation (hours), fertilizer (kilograms), and pesticide (number of times).

## 5. Empirical Results

Here, we present our estimates of the impact of treatment assignment or expanded credit access on productivity, technological change (frontier shift), and the technical efficiency of farmers. We performed the impact analysis over a 24-month period, and the results are divided into three subsections. In Subsection 5.1, we present the impact of credit access on input use and adoption of modern hybrid rice varieties. Then, in Subsection 5.2, we present the overall impact on productivity using the OLS estimation method and equation (3). We then decompose and analyze the sources of the effect on productivity, finding significant impacts of access to credit, both economically and statistically, on both productivity and efficiency. We examine the impacts relative to the amount of credit used. In Subsection 5.3, we examine the impact of access to credit broken down into various demographic and farm characteristics based on the baseline survey, and find heterogeneity of impact within the treatment group.

### 5.1 Effect of Credit Access on Input Use and Adoption of Modern Hybrid Rice Varieties

---

[33] Another reason why the sum of the two effects (from OLS) does not necessarily equal $\gamma$ estimated using the maximum likelihood method is that the maximum likelihood method uses distributional assumptions, while OLS does not.
[34] If we estimate our frontier model (equation 6 -10) with a modification of equation (10), where credit access ($Z_i$) is used as a determinant of the noise term, we find that the error/noise variance is on average higher for the households with credit access compared to the control group. And, when we compute the variance of the composed error $[v_i - u_i(Z_i)]$ as $\sigma_u^2(Z_i) + \sigma_v^2(Z_i)$, then we find it smaller for $Z_i=1$, which is primarily due to the statistically significant negative effect of $Z_i$ on inefficiency.

First, we check whether credit belongs to the production input set. The impact of access to the BCUP credit program on the use of inputs and adoption of modern hybrid rice varieties 24 months after the intervention is estimated using OLS and equation (2). The results are presented in Table 4.

[Table 4 about here]

We find that the treatment group is 15.64 percent more likely to adopt modern hybrid rice varieties than the control group. On average, treatment households use 2.26 times more pesticides, an important complementary input in modern hybrid rice production, than control households. We also find that treatment households use more land, seed, fertilizer, and machinery for land preparation but less labor and irrigation than control households. However, the standard errors relating to these variables are large, and therefore the differences are not statistically significant. Overall, the results presented in Table 4 suggest that access to credit causes a change in productivity through changes in the use of inputs and available technologies.

## 5.2     Effect of Credit on Productivity, Technological Change, and Change in Efficiency

We estimate the overall effect of being offered access to the credit program on changes in productivity 24 months after the intervention using OLS and equation (3). The estimates are presented in Table 5. We find an increase in rice yields of around 13.5 percent in treatment households compared to control households, and the impact is statistically significant at the 95 percent level. The average baseline yield of 18.12 kilograms of rice per decimal implies an increase of approximately two kilograms of rice per decimal. In Table A3 in the Appendix, we divide rice varieties into modern hybrid varieties and high-yielding varieties (HYVs) and find a statistically significant positive treatment effect for yields of both HYVs and modern hybrid varieties (around 13 and 12 percent, respectively). Overall, we find a positive effect of expanded credit access on productivity.[35]

---

[35] The findings of table 4 show that rice variety choice is an outcome of the treatment status. This endogenous selection into rice variety can be a mediating mechanism of the productivity change differences by credit availability. Therefore, the results in Table A3 cannot be interpreted as the causal effect of credit availability on the productivity of a given rice variety.

[Table 5 about here]

Table 6 shows the results from the stochastic frontier model, which decomposes the impact of credit expansion into frontier shifts and efficiency changes.[36] Columns (1) and (2) capture the effect of credit access on output that comes from a frontier shift, whereas columns (3) and (4) capture the effect that comes from efficiency improvements. We find a positive and statistically significant effect of credit access on frontier shifts. On average, around 11 percent of the overall productivity gain comes from technological change, or a frontier shift. The likely mechanism underlying this finding might be that in the absence of access to credit, households are more likely to continue with their conventional farming practices, and are unwilling to grow modern crop varieties that offer higher yields. The findings presented in Table 4 show that better access to credit enables farm households to introduce more productive modern hybrid rice varieties, which leads to a shift in the production frontier.

We obtain the impact of credit access on inefficiency after estimating the model parameters and the efficiency index (Figure 4 shows the density plot of the inefficiency index). Columns (3) and (4) of Table 6 show that small-scale farms with access to subsidized credit are, on average, 3 percent more efficient than farm households with no access to credit. Given the average baseline rice yield of 18.12 kilograms per decimal, this positive effect on efficiency implies that credit access enabled treatment households to produce approximately half a kilogram more rice per decimal than control farms as a result of improved efficiency. The positive impact of access to credit is most pronounced among producers of modern hybrid rice varieties, who exhibit efficiency gains of 9 percent on average (see Table A4 in the Appendix).

One possible explanation for these findings can be seen from the findings presented in Table 4, which show that access to credit significantly increases the adoption of hybrid rice varieties and the use of pesticides among the treatment group compared with the control group. Hybrid rice varieties offer higher potential yields than other rice varieties, but also require more complementary inputs and more timely use of variable inputs, which farmers find easier to manage with access to credit.[37] Another possible factor might be a difference in knowledge about effective farming practices and the timely use of inputs. Although our study is limited to the impact of credit rather than the combined impact of credit and extension services, our analysis shows that treatment group farmers are more likely to be familiar with and discuss crop choices, input choices, and farming practices with agricultural extension service officers and providers than control group farmers (see Table A5 in the Appendix).

---

[36] Mean baseline inefficiency is around 17 percent (estimated using equations 6–10 and baseline data), which implies that before obtaining access to credit, farmers lose around 17 percent, on average, of their potential rice output through inefficiency.
[37] It is also tempting to consider that unmeasured or poorly measured inputs will show up as efficiency. However, because of the experimental design, this potential measurement bias is likely to be the same in both the treatment and control groups, and thus will be cancelled out.

[Table 6 about here]

We also examine the impact of the amount of credit received on marginal returns while all other factors of production remain constant. Figure 4 shows our estimates of rice yields and efficiency divided into ten groups based on the amount of credit received. After taking the confidence intervals into consideration, we find that the impact on yields is uniform regardless of the amount of credit received, and thus we fail to find evidence that changes in the amount of credit received affect yields (Figure 5, panel A). In other words, regardless of the amount of credit received, the impact on marginal productivity remains the same. We also find no evidence of differences in terms of technical efficiency among farmers based on the amount of credit received (Figure 5, panel B).

[Figure 5 Panel A and Panel B about here]

### 5.3 Heterogeneous Effect of Credit Access

In this section, we explore the impact of credit access based on several demographic and farm characteristics. In particular, we focus on gender and level of education of the household head, land area, and tenancy arrangements. We augment specifications (4) and (10) to estimate the heterogeneity of the impact on output and efficiency, respectively.

To capture the heterogeneity of the effect on rice productivity, we estimate:

$$lnY_i = \gamma_0 + \gamma_1 Z_i + \sum_{k=1}^{l} \beta_k lnX_{ki} + \sum_{j=1}^{m} \theta_j I_{ji} + \sum_{j=1}^{m} \delta_j I_{ji} * Z_i + v_i, \quad (11)$$

where $I_{ij}$ is a vector of economic and demographic variables $j$ for farm household $i$. We interact

$I_{ij}$ with the household's treatment assignment status ($Z_i$). All other variables in $lnX_{ki}$ are the same as before.

Since $Z_i$ is a dummy variable from (12), the marginal effect of $Z_i$ on technological change is given by $E(Y|I,X,Z_i = 1) - E(Y|I,X,Z_i = 0) = [\gamma_0 + \gamma_1 + \sum_{k=1}^{l}\beta_k X_{ki} + \sum_{j=1}^{m}\theta_j I_{ji} + \sum_{j=1}^{m}\delta_j I_{ji}] - [\gamma_0 + \sum_{k=1}^{l}\beta_k X_{ki} + \sum_{j=1}^{m}\theta_j I_{ji}] = \gamma_1 + \sum_{j=1}^{m}\delta_j I_{ji}$. The coefficient of the interaction term $\delta_j$ in equation (12) captures the heterogeneous effect of expanded credit access within the treatment group. Note that this depends on the values of $I_j$. Our $I_j$ variables are dummy variables representing various demographic and farm characteristics, the means of which are presented in Table A6 in the Appendix.

To examine the difference between the heterogeneous and homogeneous models in terms of inefficiency, we add the inefficiency term $u_i(M_i)$ in (11) and examine the difference between the mean inefficiencies, that is, $E[u_i(M_i)|Z_i = 1] - E[u_i(M_i)|Z_i = 0]$, where $M_i$ are the determinants of inefficiency. The $M_i$ variables are the same in both the frontier function and the determinants of inefficiency. Within the treatment group, to capture the degree of heterogeneity in the effect of credit access on efficiency, we re-estimate our frontier model (equations 6–10) after modifying equation (9) as follows:

$$\sigma_u^2(M_i) = exp(\delta_0 + \delta_1 Z_i + \sum_{j=1}^{m}\rho_j I_{ji} + \sum_{j=1}^{m}\eta_j I_{ji} * Z_i), \quad (12)$$

where $M_i = \delta_0 + \delta_1 Z_i + \sum_{j=1}^{m}\rho_j I_{ji} + \sum_{j=1}^{m}\eta_j I_{ji} * Z_i$. The marginal effect of Z on mean inefficiency can be calculated as follows: $E[u_i(M_i)|Z_i = 1] - E[u_i(M_i)|Z_i = 0] = \sqrt{2/\pi} *$ .5 $(exp(\delta_0 + \delta_1 + \sum_{j=1}^{m}\rho_j I_{ji} + \sum_{j=1}^{m}\eta_j I_{ji}) - exp(\delta_0 + \sum_{j=1}^{m}\rho_j I_{ji})) = \sqrt{2/\pi} *$ (.5) $exp(\delta_1 + \sum_{j=1}^{m}\eta_j I_{ji})$. Clearly, the marginal effect of access to credit on inefficiency depends on the $I$ variables. The marginal effect via technological change is $\gamma_1 + \sum_{j=1}^{m}\delta_j I_{ji}$, which also depends on the $I$ variables.

Tables 7 and 8 show the estimates of the degree of heterogeneity in the effect of access to credit.

[Table 7 about here]
[Table 8 about here]

Columns 1–4 of Tables 7 and 8 show the effects of credit access based on the gender and level of education of the household head. Several studies have found gender differences in the take-up of credit, use of fertilizer, use of capital, and adoption of new technology (Udry 1996, Tiruneh et al. 2001). Belanger and Li (2009) find that women have less control over assets, access to credit, and influence in decision-making regarding extension services and inputs, resulting in lower farmer productivity. We found that female-headed farms that are provided with access to credit generate, on average, approximately 7 percent more in terms of output than male-led farms with credit access (see Table 7). In terms of efficiency (Table 8), we found that female-led enterprises with access to credit were 1.5 percent more efficient than male-led enterprises with access to credit. However, the results were not statistically significant. Our findings in relation to the education level of the household head were similar, but once again, not statistically significant.

Next, we consider the baseline farm size. Previous empirical studies have found an inverse relationship between farm size and output per hectare (Cornia 1985; Fan and Chan-Kang 2003). Some studies have suggested that this is the result of errors in measuring soil quality and land size (Fan and Chan-Kang 2003), while other studies have found that this inverse relationship disappears at high levels of technology adoption (Cornia 1985). We examined the relationship between credit access and yield or efficiency based on farm land size and found a negative relationship, suggesting that within the treatment group, the average effect of access to credit is greater for larger farms. However, we did not find a statistically significant difference between large and small farms in terms of estimates of heterogeneity in relation to the effect of access to credit.

Finally, we tested for differences in the impact of access to credit based on land ownership and tenancy status. For both technological change and efficiency outcomes, we found a significantly positive effect for pure tenant and mixed-tenant farm enterprises compared with farm enterprises that only cultivated their own land. The marginal effect of credit access on productivity was around 14 percent for pure tenant farm households (i.e., those that only cultivated other people's land).[38] Columns 7 and 8 of Table 7 show that within the treatment group, the effect of credit on productivity is approximately 3.5 percent higher for tenant farm households than for farmers who cultivate their own land. In the case of efficiency change, the impact for pure tenant farms is, on average, 5 percent higher than for owner farms (column 8 of Table 8). In terms of farming practices, we found that when there is access to credit, adoption of hybrid rice varieties is significantly higher among tenant farm enterprises than among farms that cultivate their own land. This suggests that relatively resource-poor farm enterprises gain more from access to credit.

---

[38] From columns 7 and 8, it can be seen that $\gamma_1$= 13.06, $\delta_j$=3.45. The mean for pure tenant farm households is 0.32, which implies that 32 percent of farm households in the sample only cultivate other people's land, therefore the effect of the treatment assignment is (13.06+(3.45*0.32)) =14.16.

## 6. Discussion and Concluding Remarks

Access to subsidized credit can aid small farm households in increasing their productivity by enabling them to adopt better technology and/or enabling efficiency improvements. In this study, we analyze data from a field experiment based on the random assignment of credit access in Bangladesh to estimate the impact of credit expansion on farm productivity. In particular, we examine whether the productivity increase is the result of changes in technology or improved efficiency. First, we examine whether being offered access to the credit program changes the amounts and types of inputs used. Then, we estimate the average overall impact of credit access on rice yields and examine the sources of changes in productivity. We use the stochastic production frontier model as a tool to disentangle the two effects, technological change and change in efficiency.

We find that relaxing the credit constraint has a significant positive impact on rice yields, via both technological change and improved efficiency of farmers. We find a positive impact of access to credit on total productivity that is statistically significant at the 1 percent level. On average, we find a productivity increase of around 14 percent among farmers provided with access to credit services. After decomposing the overall output effect into frontier shift and efficiency change effects, we find that most of the effect, around 11 percent, is related to a frontier shift, that is, the adoption of modern hybrid rice varieties and the use of complementary inputs. In terms of technical efficiency, we find that small-scale farms with access to subsidized credit experience, on average, a 3 percent increase in efficiency compared with households with no access to credit. This effect is even more pronounced in relation to modern hybrid rice varieties, which deliver efficiency gains of around 9 percent on average. We find no evidence of more sizable impacts on yields and efficiency among farmers that take up larger amounts of credit. Within the treatment group, the impact is greater among pure tenant and mixed-tenant farming households than among farmers that only cultivate their own land.

A simple story helps to explain the positive impacts that we observed. When farmers have limited recourse to well-functioning credit markets, they are unlikely to adopt modern high-yielding crop varieties that require more cash upfront to buy seed and complementary inputs that must be obtained and used in a timely manner. Provision of credit provides a liquidity buffer that enables these farmers to adopt modern crop varieties and apply and manage complementary inputs in a more effective and timely manner, which ultimately leads to higher productivity and efficiency compared with households that do not have access to credit. We find that on average, households with access to BCUP credit are more likely to adopt modern hybrid rice varieties than households with no access to credit. Pesticides are essential in the production process and for the stability of yields of hybrid rice, and we find that households in the treatment group procure significantly more pesticides than those in the control group. When credit is available, adoption of hybrid rice varieties is significantly higher among tenant farm enterprises than among enterprises that cultivate their own land, which suggests that the more resource-poor the farm enterprise, the greater the benefit from obtaining access to credit.

The findings of this study have important implications for policy, especially in relation to resource-constrained contexts. This study adds to our knowledge of the potential benefits of credit programs targeting subsistence farm enterprises, and the findings can help inform decisions aimed at achieving better targeting by such programs.

**Figure 1: Design of Field Experiment and Treatment Assignment**

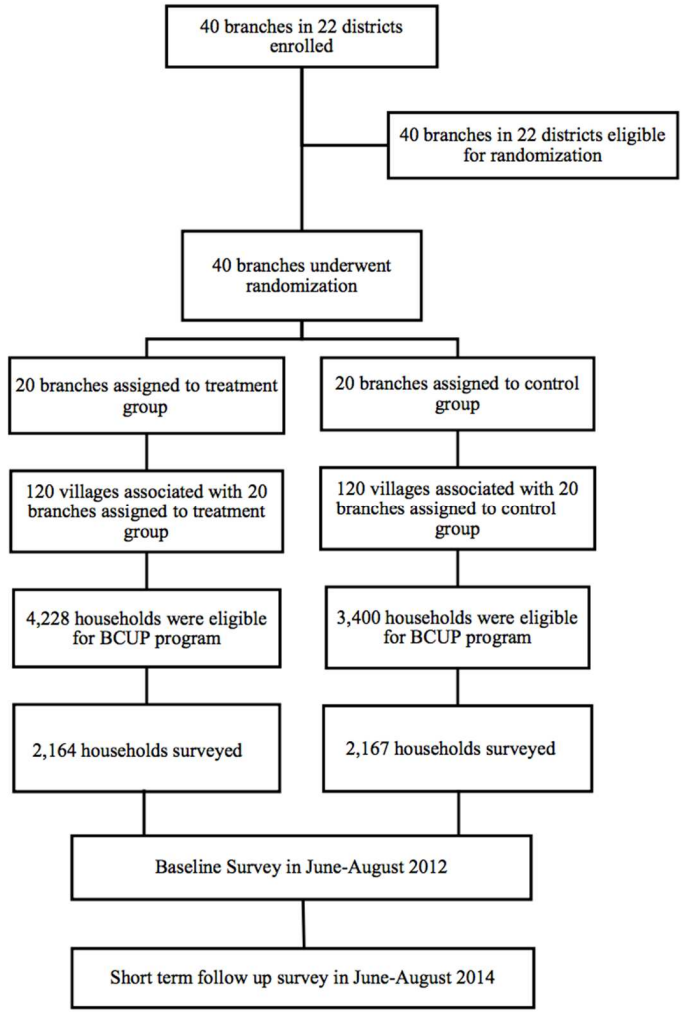

**Figure 2: Map of Treatment and Control Areas**

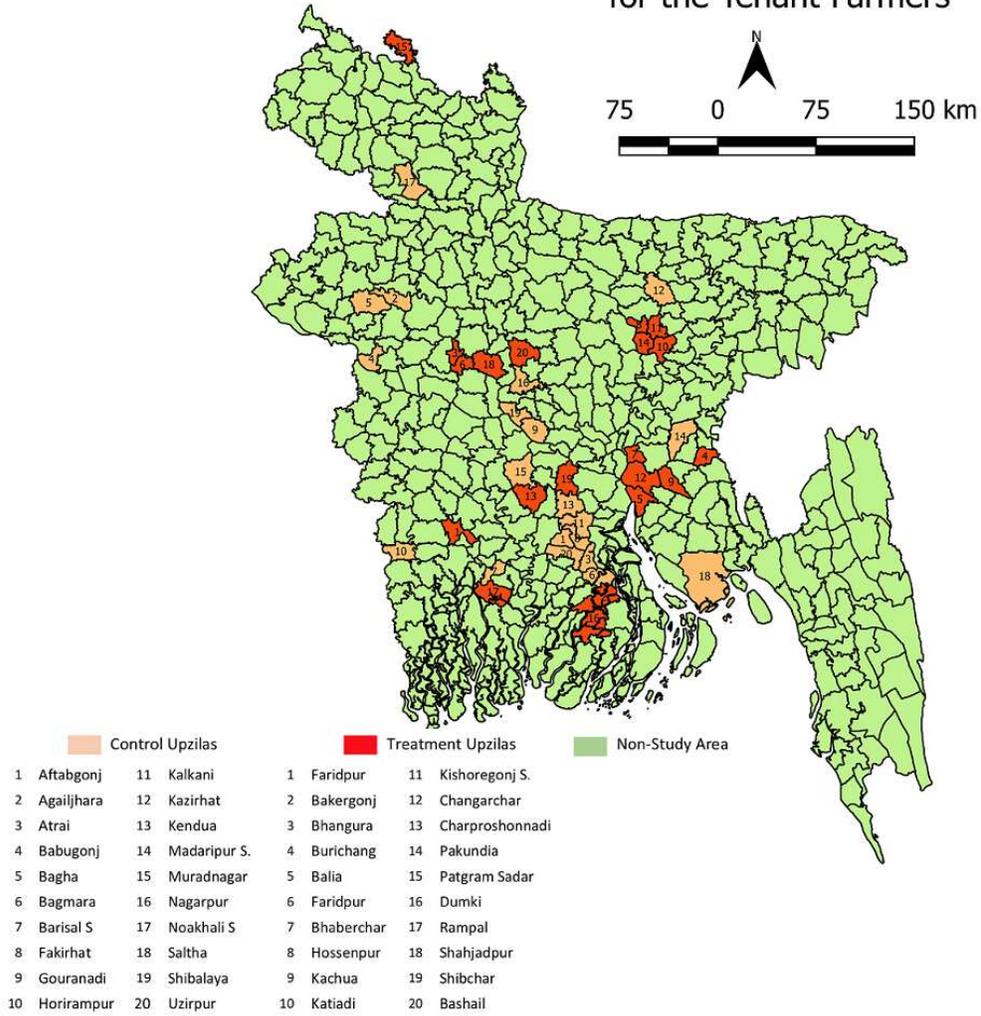

**Figure 3: BCUP Program Features**

| Treatment Groups | Program Features |
|---|---|
| Treatment Group | Credit Limit: 5,000 taka-120,000 taka[*]<br>Duration: 6-10 months<br>Grace Period: 1 month<br>Installment: monthly<br>Interest Rate: 10percent (flat)[**] |
| Control Group | None |

*Note:* [*]79 taka=1 USD; [**]In the flat rate method, interest is charged on the full original loan amount throughout the loan term whereas in the declining balance method, interest calculation is based on the outstanding loan balance – the balance of money that remains in the borrower's hands as the loan is repaid during the loan term. BCUP provided loans to farmers at subsided interest rate of 10 per cent per year (flat rate). The effective rate of interest is about 15 to 20 per cent on declining balance method, depending on the mode of repayment of the principal and interest due. As per the rules of the Microcredit Regulatory (MRA) of the Bangladesh Bank, NGOs can charge up to a maximum of 27 per cent rate of interest on declining balance for their microfinance operations.

**Figure 4: Density Plot of Inefficiency Index**

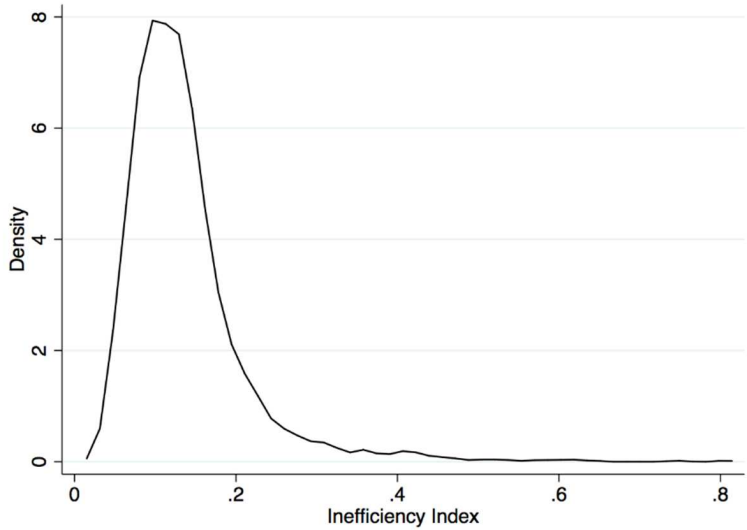

# Figure 5: Impact of Credit by the Amount of Credit Taken

**Panel A: Impact on Rice Yield**

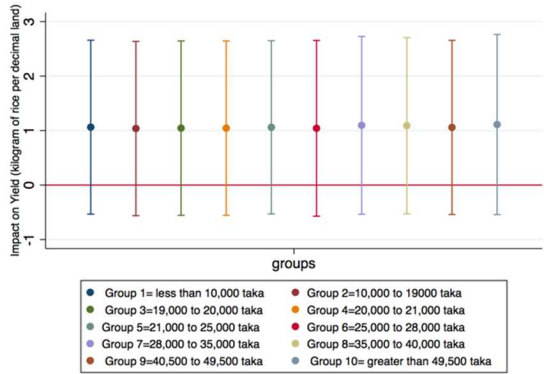

**Note**: 79 taka= 1 USD

**Panel B: Impact on Efficiency (percentage effect)**

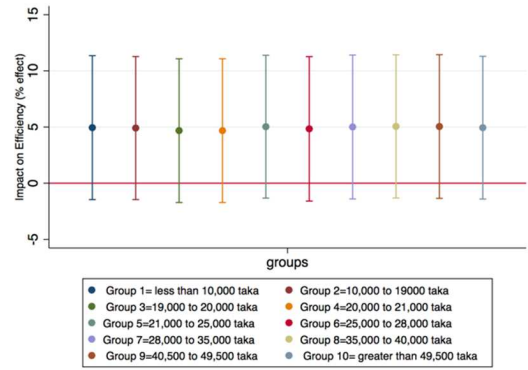

**Note**: 79 taka= 1 USD

## Table 1: Descriptive Statistics and Baseline Characteristics

| Variables | Observations | Mean | Std. Dev. |
|---|---|---|---|
| **Household Composition** | | | |
| Female Headed household | 3,292 | 0.07 | 0.26 |
| Age of household head (in years) | 3,292 | 44.82 | 11.66 |
| Household size (number of members) | 3,292 | 4.87 | 1.79 |
| Number of children (<16 years) | 3,292 | 3.11 | 1.35 |
| Number of adult members (>16 years) | 3,292 | 1.75 | 1.25 |
| Household head with no education | 3,292 | 0.45 | 0.49 |
| **Amount of Land** | | | |
| Own cultivated land (in decimal[1]) | 3,292 | 39.38 | 51.75 |
| Rented in land (in decimal) | 3,292 | 48.63 | 64.57 |
| Rented out land (in decimal) | 3,292 | 8.08 | 27.52 |
| Total cultivated land (in decimal) | 3,292 | 88.02 | 75.17 |
| **Amount of Credit and Interest Rate** | | | |
| Formal and informal loan amount (in taka) | 3,292 | 5221.04 | 28101.67 |
| Interest rate of loans from formal[2] institutions (percent) | 320 | 11.15 | 3.91 |
| Interest rate of loans from informal[3] institutions (percent) | 287 | 18.00 | 12.83 |
| **Output and Input for Rice Production (yearly)** | | | |
| Rice yield (total rice/total land)[4] | 3,292 | 18.12 | 4.27 |
| High Yielding Variety rice yield (HYV rice/ land in HYV rice) | 3,218 | 17.02 | 4.17 |
| Hybrid rice yield (HB rice/ land in HB rice) | 213 | 20.18 | 7.18 |
| Traditional Variety rice yield (TV rice/ land in TV rice) | 1,307 | 9.83 | 3.44 |
| Total land (in decimal) | 3,292 | 99.70 | 85.28 |
| Total labor days (own as well as hired labor) | 3,292 | 46.37 | 39.80 |
| Total plough (number of times) | 3,292 | 6.61 | 6.03 |
| Total seed (kilogram) | 3,292 | 17.27 | 18.80 |
| Total irrigation (hours) | 3,292 | 45.60 | 51.62 |
| Total fertilizer (kilogram) | 3,292 | 168.30 | 150.62 |
| Total pesticide (number of times) | 3,292 | 4.27 | 5.65 |

*Notes*: **Unit of observation**: Household. Sample includes all rice producing farm households surveyed at baseline (2012). [1]Land is measured in decimal. A **decimal** (also spelled **decimel**) is a unit of area in India and Bangladesh approximately equal to 1/100 acre (40.46 m²); 247 decimal=1 hectare. [2]**Formal** institutions include bank and cooperatives. [3]**Informal lenders** include moneylenders, loans from friends/family, and buying goods/services on credit from sellers. [4]Rice in measured in kilogram (1 kilogram=2.204 pounds).
**Own land** refers to the cultivated crop land owned by the farm household. **Rented in** land means the land rented in from others for crop cultivation. **Rented out** land means the land rented to other farmers for crop cultivation.
**High Yielding Varieties (HYV)** rice seeds are land substituting, water economizing, more labor using, and employment generating innovation. HYVs significantly outperform traditional varieties in the presence of an efficient management of irrigation, pesticides, and fertilizers. However, in the absence of these inputs, traditional varieties may outperform HYVs.
**Hybrid rice** is any genealogy of rice produced by crossbreeding different kinds of rice. It typically displays heterosis or hybrid vigor such that when it is grown under the same conditions as comparable high yielding inbred rice varieties it can produce up to 30percent more rice. However, the heterosis effect disappears after the first (F1) generation, so the farmers cannot save seeds produced from a hybrid crop and need to purchase new F1 seeds every season to get the heterosis effect each time.

## Table 2: Baseline Characteristics and Balancing

|  | Means by treatment | | |
|---|---|---|---|
|  | Control (1) | Treatment (2) | $P$-value[1] (3) |
| **Variables** | | | |
| **Household Composition** | | | |
| Female headed household (percentage) | 0.05 (0.01) | 0.09 (0.01) | 0.02 |
| Age of household head (in years) | 44.67 (0.28) | 44.98 (0.26) | 0.69 |
| Household size (number of members) | 4.82 (0.04) | 4.92 (0.05) | 0.59 |
| Number of adult members (>16 years) | 3.15 (0.03) | 3.08 (0.03) | 0.49 |
| Number of children (<16 years) | 1.67 (0.03) | 1.84 (0.03) | 0.19 |
| Household head with no education | 0.45 (0.01) | 0.45 (0.01) | 0.92 |
| **Amount of Land** | | | |
| Own cultivated land (in decimal[2]) | 40.92 (1.29) | 37.76 (1.25) | 0.35 |
| Rented in land (in decimal) | 49.86 (1.69) | 47.32 (1.48) | 0.61 |
| Rented out land (in decimal) | 8.24 (0.67) | 7.92 (0.69) | 0.78 |
| Total cultivated land (in decimal) | 90.79 (1.96) | 85.09 (1.71) | 0.42 |
| **Amount of Credit and Interest Rate** | | | |
| Formal and informal[3] loan amount (in taka) | 5136.45 (581.53) | 5,316.45 (798.93) | 0.91 |
| Interest rate of loans from formal institutions (percent) | 10.43 (3.55) | 11.24 (4.07) | 0.39 |
| Interest rate of loans from informal institutions (percent) | 18.00 (15.36) | 18.47 (15.50) | 0.89 |
| **Access to Other BRAC Programs and Services** | | | |
| Member of other BRAC loan programs (dummy) | 0.02 (0.00) | 0.01 (0.00) | 0.76 |
| Received other services besides credit (dummy) | 0.01 (0.00) | 0.00 (0.00) | 0.51 |

*Notes*: [1]**Column 3** shows the P value of the test of equality of means by random assignment of credit access (whether the means of the variables are statistically significantly similar to each other). [2]A **decimal** (also spelled **decimel**) is a unit of area in India and Bangladesh approximately equal to 1/100 acre (40.46 m²); 247 decimals=1 hectare. [3]The formal source includes government bank, commercial bank, and other government and non-government loan institutions. Informal sources include friends and relatives, traditional moneylenders, landlords etc.
**Unit of observation**: Household. Sample includes all rice producing farm households surveyed at baseline 2012 (3,292 households). Standard errors (in parentheses) are clustered at Branch level. Informal lenders include moneylenders, loans from friends/family, and buying goods/services on credit from sellers. **Own land** refers to the cultivated crop land owned by the farm household. **Rented in** land means the land rented in from others for crop cultivation. **Rented out** land means the land rented to other farmers for crop cultivation.

## Table 2 (contd.): Baseline Characteristics and Balancing

|  | Means by treatment | | |
|---|---|---|---|
|  | Control (1) | Treatment (2) | P-value (3) |
| **Variables** | | | |
| **Output and Input for Rice Production (yearly)** | | | |
| Rice yield (total rice/total land)[4] | 18.79 | 17.40 | 0.24 |
|  | (0.12) | (0.10) |  |
| High Yielding Variety rice yield (HYV rice/ land in HYV rice) | 18.66 | 17.33 | 0.28 |
|  | (0.11) | (0.09) |  |
| Hybrid rice yield (HB rice/ land in HB rice) | 20.39 | 19.94 | 0.79 |
|  | (0.73) | (0.65) |  |
| Traditional Variety rice yield (TV rice/ land in TV rice) | 9.45 | 10.23 | 0.98 |
|  | (0.13) | (0.14) |  |
| Total Land (in decimal) | 91.39 | 108.52 | 0.10 |
|  | (1.85) | (2.12) |  |
| Total Labor (days) | 41.99 | 51.03 | 0.10 |
|  | (0.87) | (1.08) |  |
| Total Plough (number of times) | 6.40 | 6.85 | 0.48 |
|  | (0.14) | (0.14) |  |
| Total Seed (kilogram) | 17.36 | 17.18 | 0.94 |
|  | (0.49) | (0.43) |  |
| Total Irrigation (hours) | 44.27 | 47.03 | 0.74 |
|  | (1.19) | (1.15) |  |
| Total Fertilizer (kilogram) | 157.79 | 179.45 | 0.11 |
|  | (3.42) | (3.57) |  |
| Total Pesticide (number of times) | 4.58 | 3.96 | 0.38 |
|  | (0.14) | (0.11) |  |
| Observations | 1,694 | 1,598 |  |
| **Joint significance test** |  | $F(16, 39) =$ | 1.64 |
|  |  | $Prob > F =$ | 0.13 |

*Notes*: [4]Rice in measured in kilogram (1 kilogram=2.204 pounds). Land is measured in decimal (247 decimals=1 hectare.) [††]**kilogram** (1 kilogram=2.204 pounds). **Column 3** shows the P value of the test of equality of means by random assignment of credit access (whether the means of the variables are statistically significantly similar to each other).
**Unit of observation**: Household. Sample includes all rice producing farm households surveyed at baseline 2012 (3,292 households). Standard errors (in parentheses) are clustered at Branch level. **High Yielding Varieties (HYV)** rice seeds are land substituting, water economizing, more labor using, and employment generating innovation. HYVs significantly outperform traditional varieties in the presence of an efficient management of irrigation, pesticides, and fertilizers. However, in the absence of these inputs, traditional varieties may outperform HYVs.
**Hybrid rice** is any genealogy of rice produced by crossbreeding different kinds of rice. It typically displays heterosis or hybrid vigor such that when it is grown under the same conditions as comparable high yielding inbred rice varieties it can produce up to 30percent more rice. However, the heterosis effect disappears after the first (F1) generation, so the farmers cannot save seeds produced from a hybrid crop and need to purchase new F1 seeds every season to get the heterosis effect each time.

## Table 3: Attrition Rate

Dependent Variable: Attrited Households (1= household with baseline information but no follow-up information)

|  | (1) | (2) |
|---|---|---|
| Treatment Assignment (1=household assigned to treatment group) | -0.02 | -0.03 |
| Female Headed household (percentage) |  | 0.16** |
| Age of household head (in years) |  | 0.00 |
| Household head with no education (percentage) |  | -0.02* |
| Any baseline formal and informal loan (yes=1) |  | 0.01 |
| Observations | 3,755 | 3,755 |

*Notes:* Unit of observation: Household. Sample includes all rice producing farm households surveyed at baseline (2012). Informal lenders include moneylenders, loans from friends/family, and buying goods/services on credit from sellers.

## Table 4: Impact of Access to Credit on Input Use and Adoption of Modern Hybrid Rice

| Variables | Effect of Credit Access | Observations |
|---|---|---|
| Total Land (in decimal) | 2.03 | 3,172 |
|  | (14.68) |  |
| Total Labor (days) | -0.83 | 3,172 |
|  | (7.88) |  |
| Total Seed (kilogram) | 4.15 | 3,172 |
|  | (3.27) |  |
| Total Irrigation (hours) | -3.36 | 3,172 |
|  | (12.24) |  |
| Total Fertilizer (kilogram) | 27.19 | 3,172 |
|  | (29.73) |  |
| Total Pesticide (number of times) | 2.26** | 3,172 |
|  | (1.04) |  |
| Total Plough (number of times) | 0.95 | 3,172 |
|  | (0.96) |  |
| Adoption of Modern Hybrid Rice[1] (dummy) | 14.38*** | 3,172 |
|  | (2.50) |  |
| Control for Baseline Covariates | Yes |  |

*Notes*: ***$p<0.01$, **$p<0.05$, *$p<0.1$. [†]Rice is measured in kilogram (1 kilogram=2.204 pounds). Land is measured in decimal (also spelled **decimal**) which is a unit of area in India and Bangladesh approximately equal to 1/100 acre (40.46 m²); 247 decimals=1 hectares. Standard errors (in parentheses) are clustered at the branch level. [1]Adoption is a dummy variable that takes a value of 1 if the farm produces Hybrid rice in endline but has zero baseline production.

**Table 5: Impact of Access to Credit on Productivity of Rice**

| Variables | Effect of Credit on Productivity (1) | Effect of Credit on Productivity (2) | Observations |
|---|---|---|---|
| Rice yield (Total rice/Total land)[1] | 13.54*** | 13.80*** | 2,267 |
|  | (3.07) | (3.03) |  |
| Average rice yield at baseline (Total rice/Total land) | 18.12 | | |
| Model includes all other production inputs | yes | yes | |
| Control for Baseline covariates | no | yes | |

*Notes*: *** p<0.01, ** p<0.05, * p<0.1. [1]Rice is measured in kilogram (1 kilogram=2.204 pounds). Land is measured in decimal (also spelled **decimel**) which is a unit of area in India and Bangladesh approximately equal to 1/100 acre (40.46 m²); 247 decimals=1 hectares. Column (1) and (2) shows the impact of credit access on outcome of interest. Sample includes rice producing farm households. Standard errors (in parentheses) are clustered at the Branch level.

**Table 6: Impact of Access to Credit on Frontier Shift and Efficiency of Rice Production (percentage effect)**

| Variables | Dependent Variable: Rice Yield (kilogram of rice per decimal of land) | | | |
|---|---|---|---|---|
|  | Frontier Shift | Frontier Shift | Inefficiency | Inefficiency |
|  | (1) | (2) | (3) | (4) |
| Credit access (1=assigned in treatment group) | 10.67*** | 10.79*** | -2.97* | -3.18** |
|  | (1.15) | (1.11) | (1.57) | (1.42) |
| Mean Baseline Inefficiency in Rice Production | | 17.15 | | |
| Model includes all other production inputs | yes | yes | yes | yes |
| Baseline Covariates | no | yes | no | yes |
| Observations | 2,267 | 2,267 | 2,267 | 2,267 |

*Notes*: Unit of observation is household. Sample includes rice producing farm households. Rice in measured in kilogram (1 kilogram=2.204 pounds). Land is measured in decimal which is a unit of area in India and Bangladesh approximately equal to 1/100 acre (40.46 m2); 247 decimals=1 hectare. Standard error in parenthesis and are clustered at branch level. *** p<0.01, ** p<0.05, * p<0.1

## Table 7: Heterogeneous Impact of Access to Credit on Rice Production

| Variables | Dependent Variable: Rice Yield (kilogram of rice per decimal of land) | | | | | | | |
|---|---|---|---|---|---|---|---|---|
| | (1) | (2) | (3) | (4) | (5) | (6) | (7) | (8) |
| Credit access (1=assigned in treatment group) | 13.31*** | 13.40*** | 13.31*** | 13.45*** | 14.27*** | 14.36*** | 12.96*** | 13.06*** |
| | (3.04) | (3.09) | (3.23) | (3.20) | (3.04) | (3.00) | (3.01) | (2.99) |
| Female headed household (dummy) | -6.28 | -6.42 | | | | | | |
| | (4.81) | (4.75) | | | | | | |
| Access to credit*Female headed household | 6.69 | 7.13 | | | | | | |
| | (4.87) | (4.82) | | | | | | |
| Head with no education (dummy) | | | 1.23 | 1.21 | | | | |
| | | | (1.30) | (1.41) | | | | |
| Access to credit*Head with no education | | | 0.47 | 0.45 | | | | |
| | | | (1.30) | (0.62) | | | | |
| Small farm size (1=cultivated land<50 decimal) | | | | | 1.73 | 1.65 | | |
| | | | | | (1.58) | (1.55) | | |
| Access to credit*Small farm size (1=cultivated land<50 decimal) | | | | | -2.82 | -0.27 | | |
| | | | | | (1.87) | (1.88) | | |
| Model includes all other production inputs | yes | yes | yes | yes | yes | yes | yes | yes |
| Control for Baseline Covariates | no | yes | no | yes | no | yes | no | yes |
| Mean Baseline Rice Yield (Total rice/Total land) | | | | 18.12 | | | | |
| Observations | 2,267 | 2,267 | 2,267 | 2,267 | 2,267 | 2,267 | 2,267 | 2,267 |

*Notes*: *** p<0.01, ** p<0.05, * p<0.1. Sample includes all rice producing farm households. Rice is measured in kilogram (1 kilogram=2.204 pounds). Land is measured in decimal (also spelled **decimel**) which is a unit of area in India and Bangladesh approximately equal to 1/100 acre (40.46 m²); 247 decimals=1 hectare. Results show the percentage effect of access to credit on rice production efficiency by different farm household characteristics. Small farm size takes a value 1 if total cultivated land by farm household is less than 50 decimals. Familiarity with agriculture extension service provider implies that farmers are acquainted with the persons/institution from whom they can seek information or advice on crop selection, crop rotation, modern cropping technology, appropriate use of fertilizer, pesticide etc. Standard errors (in parentheses) are clustered at Branch level.

## Table 7 (contd.): Heterogeneous Impact of Access to Credit on Rice Production

| Variables | Dependent Variable: Rice Yield (kilogram of rice per decimal of land) | | | | | | | |
|---|---|---|---|---|---|---|---|---|
| | (1) | (2) | (3) | (4) | (5) | (6) | (7) | (8) |
| Pure tenant farm households (1= no own land rice cultivation) | | | | | | | -2.10* | -1.53 |
| | | | | | | | (1.13) | (1.40) |
| Access to credit*Pure tenant farm households (1= no own land rice cultivation) | | | | | | | 3.50** | 3.45** |
| | | | | | | | (1.54) | (1.45) |
| Mixed tenant farms (1=cultivate own land as well as others land) | | | | | | | 0.45 | 0.58 |
| | | | | | | | (1.18) | (1.43) |
| Access to credit*Mixed tenant farms (1=cultivate own land as well as others land) | | | | | | | 1.79 | 1.74 |
| | | | | | | | (1.55) | (1.56) |
| Model includes all other production inputs | yes | yes | yes | yes | yes | yes | yes | yes |
| Control for Baseline Covariates | no | yes | no | yes | no | yes | no | yes |
| Mean Baseline Rice Yield (Total rice/Total land) | | | | 18.12 | | | | |
| Observations | 2,267 | 2,267 | 2,267 | 2,267 | 2,267 | 2,267 | 2,267 | 2,267 |

*Notes*: *** $p<0.01$, ** $p<0.05$, * $p<0.1$. Sample includes all rice producing farm households. †Rice is measured in kilogram (1 kilogram=2.204 pounds). Land is measured in decimal (also spelled **decimel**) which is a unit of area in India and Bangladesh approximately equal to 1/100 acre (40.46 m²); 247 decimals=1 hectare. Results show the percentage effect of access to credit on rice production efficiency by different farm household characteristics. Pure owner farm households only cultivated their own land. Pure tenant farm households have no rice cultivation in own land- they chose either share cropping or rent or both. Standard errors (in parentheses) are clustered at Branch level.

Table 8: Heterogeneous Impact of Access to Credit on Efficiency of Rice Production (percentage effect)

| Variables | Dependent Variable: Inefficiency in Total Rice Yield | | | | | | | |
|---|---|---|---|---|---|---|---|---|
| | (1) | (2) | (3) | (4) | (5) | (6) | (7) | (8) |
| Credit access (1=assigned in treatment group) | -2.41 | -2.41* | -2.7* | -2.7* | -4.03** | -4.03** | -2.38 | -2.68* |
| | (1.50) | (1.39) | (1.56) | (1.55) | (1.55) | (1.44) | (1.47) | (1.54) |
| Female headed household (dummy) | 5.85** | 0.99 | | | | | | |
| | (2.61) | (2.72) | | | | | | |
| Access to credit*Female headed household | -8.44** | -1.49 | | | | | | |
| | (3.08) | (3.19) | | | | | | |
| Head with no education (dummy) | | | -1.23 | -1.30 | | | | |
| | | | (1.03) | (1.03) | | | | |
| Access to credit*Head with no education | | | -2.04 | -1.69 | | | | |
| | | | (1.5) | (7.24) | | | | |
| Small farm size (1=cultivated land<50 decimal) | | | | | -0.07 | -0.44 | | |
| | | | | | (1.25) | (1.22) | | |
| Access to credit*Small farm size (1=cultivated land<50 decimal) | | | | | 3.10 | 2.98 | | |
| | | | | | (1.87) | (1.80) | | |
| Model includes all other production inputs | yes | yes | yes | yes | yes | yes | yes | yes |
| Control for Baseline Covariates | no | yes | no | yes | no | yes | no | yes |
| Mean Baseline Inefficiency in Rice Production | | | | 17.15 | | | | |
| Observations | 2,267 | 2,267 | 2,267 | 2,267 | 2,267 | 2,267 | 2,267 | 2,267 |

*Notes*: Sample includes all rice producing farm households. Rice is measured in kilogram (1 kilogram=2.204 pounds). Land is measured in decimal (also spelled **decimal**) which is a unit of area in India and Bangladesh approximately equal to 1/100 acre (40.46 m2); 247 decimals=1 hectare. Results show the percentage effect of access to credit on rice production efficiency by different farm household characteristics. Small farm size takes a value 1 if total cultivated land by farm household is less than 50 decimals. Standard errors (in parentheses) are clustered at Branch level.

**Table 8 (contd.): Heterogeneous Impact of Access to Credit on Efficiency of Rice Production**
**(percentage effect)**

| Variables | Dependent Variable: Inefficiency in Total Rice Yield | | | | | | | |
|---|---|---|---|---|---|---|---|---|
| | (1) | (2) | (3) | (4) | (5) | (6) | (7) | (8) |
| Pure tenant farm households (1= no own land rice cultivation) | | | | | | | 2.41** | 1.15 |
| | | | | | | | (1.11) | (1.11) |
| Access to credit*Pure tenant farm households (1= no own land rice cultivation) | | | | | | | -5.05** | -4.77*** |
| | | | | | | | (1.83) | (1.54) |
| Mixed tenant farms (1=cultivate own land as well as others land) | | | | | | | 0.34 | -1.54 |
| | | | | | | | (1.32) | (1.32) |
| Access to credit*Mixed tenant farms (1=cultivate own land as well as others land) | | | | | | | -6.20** | -2.42** |
| | | | | | | | (2.19) | (0.76) |
| Model includes all other production inputs | yes | yes | yes | yes | yes | yes | yes | yes |
| Control for Baseline Covariates | no | yes | no | yes | no | yes | no | yes |
| Mean Baseline Inefficiency in Rice Production | | | | 17.15 | | | | |
| Observations | 2,267 | 2,267 | 2,267 | 2,267 | 2,267 | 2,267 | 2,267 | 2,267 |

*Notes*: Sample includes all rice producing farm households. Rice is measured in kilogram (1 kilogram=2.204 pounds). Land is measured in decimal (also spelled decimal) which is a unit of area in India and Bangladesh approximately equal to 1/100 acre (40.46 m2); 247 decimals=1 hectare. Results show the percentage effect of access to credit on rice production efficiency by different farm household characteristics. Pure tenant farm households have no rice cultivation in own land- they chose either share cropping or rent or both. Mixed tenants cultivate own as well as others land. Pure and Mixed tenants are compared to the base category of Pure owner farm households who only cultivated their own land. Standard errors (in parentheses) are clustered at Branch level.

# Appendix A

**Figure A1: GIS mapping for southern Region under study areas**

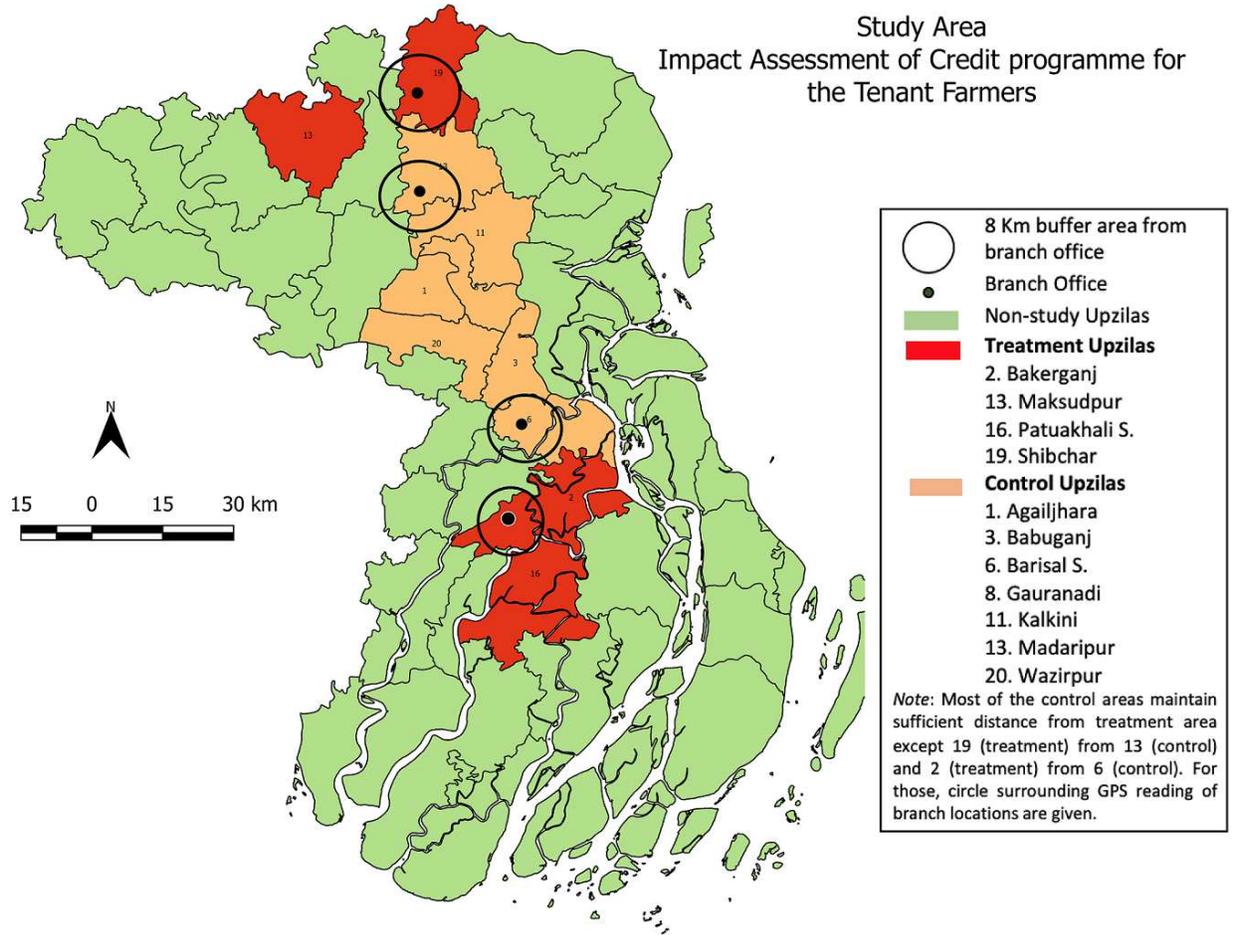

**Table A1: Baseline Characteristics and Balancing for HYV Rice Producing Households**

|  | Means by treatment | | |
|---|---|---|---|
|  | Control (1) | Treatment (2) | P-value (3) |
| **Variables** | | | |
| **Household Composition** | | | |
| Female headed household | 0.05 | 0.09 | 0.02 |
|  | (0.01) | (0.01) |  |
| Age of household head (in years) | 44.58 | 44.99 | 0.59 |
|  | (0.29) | (0.29) |  |
| Household size (number of members) | 4.83 | 4.93 | 0.6 |
|  | (0.04) | (0.05) |  |
| Number of adult members (>16 years) | 3.15 | 3.09 | 0.55 |
|  | (0.03) | (0.04) |  |
| Number of child (<16 years) | 1.68 | 1.84 | 0.22 |
|  | (0.03) | (0.03) |  |
| Household head with no education | 0.45 | 0.46 | 0.87 |
|  | (0.01) | (0.01) |  |
| **Amount of Land and Credit** | | | |
| Own cultivated land (in decimal†) | 40.96 | 37.8 | 0.35 |
|  | (1.31) | (1.27) |  |
| Rented in land (in decimal) | 50.04 | 46.82 | 0.51 |
|  | (1.71) | (1.40) |  |
| Rented out land (in decimal) | 8.40 | 7.97 | 0.71 |
|  | (0.69) | (0.70) |  |
| Total cultivated land (in decimal) | 91.00 | 84.62 | 0.37 |
|  | (2.00) | (1.64) |  |
| Formal and informal loan amount (in taka) | 5084.18 | 5304.47 | 0.89 |
|  | (591.44) | (810.23) |  |

*Notes*: †A **decimal** (also spelled **decimel**) is a unit of area in India and Bangladesh approximately equal to 1/100 acre (40.46 m²); 247 decimals=1 hectares.   ††**kilogram** (1 kilogram=2.204 pounds). Column 3 shows the P value of mean difference column 3=column1- column2.
**Unit of observation**: Household. Sample includes all rice producing farm households surveyed at baseline (2012). Standard errors (in parentheses) are clustered at Branch level. Informal lenders include moneylenders, loans from friends/family, and buying goods/services on credit from sellers.
**Own land** refers to the cultivated crop land owned by the farm household. **Rented in** land means the land rented in from others for crop cultivation. **Rented out** land means the land rented to other farmers for crop cultivation.

## Table A1 (contd.): Baseline Characteristics and Balancing for HYV Rice Producing Households

| | Means by treatment | | |
|---|---|---|---|
| | Control (1) | Treatment (2) | P-value (3) |
| **Variables** | | | |
| **Output and Input for Rice Production (yearly)** | | | |
| Rice yield (total rice/total land) †† | 17.52 | 16.63 | 0.21 |
| | (0.11) | (0.09) | |
| High Yielding Variety rice yield (HYV rice/ land in HYV rice) | 18.66 | 17.33 | 0.07 |
| | (0.11) | (0.09) | |
| Hybrid rice yield (HB rice/ land in HB rice) | 19.09 | 21.73 | 0.17 |
| | (0.9) | (0.66) | |
| Traditional Variety rice yield (TV rice/ land in TV rice) | 9.45 | 9.96 | 0.53 |
| | (0.16) | (0.20) | |
| Total Land (in decimal) | 115.38 | 123.28 | 0.55 |
| | (2.58) | (2.50) | |
| Total Labor (days) | 47.69 | 55.44 | 0.07 |
| | (1.04) | (1.15) | |
| Total Plough (number of times) | 7.26 | 7.47 | 0.77 |
| | (0.17) | (0.16) | |
| Total Seed (kilogram) | 20.82 | 19.34 | 0.57 |
| | (0.56) | (0.45) | |
| Total Irrigation (hours) | 47.92 | 47.45 | 0.96 |
| | (1.35) | (1.30) | |
| Total Fertilizer (kilogram) | 174.78 | 189.24 | 0.36 |
| | (4.1) | (3.97) | |
| Total Pesticide (number of times) | 4.97 | 4.14 | 0.27 |
| | (0.16) | (0.13) | |

*Notes*: †† Rice in measured in kilogram (1 kilogram=2.204 pounds). Land is measured in decimal (247 decimals=1 hectares.) Column 3 shows the P value of mean difference column 3=column1- column2.

**Unit of observation**: Household. Sample includes all rice producing farm households surveyed at baseline (2012). Standard errors (in parentheses) are clustered at Branch level. **High Yielding Varieties (HYV)** rice seeds are land substituting, water economizing, more labor using, and employment generating innovation. HYVs significantly outperform traditional varieties in the presence of an efficient management of irrigation, pesticides, and fertilizers. However, in the absence of these inputs, traditional varieties may outperform HYVs.

**Hybrid rice** is any genealogy of rice produced by crossbreeding different kinds of rice. It typically displays heterosis or hybrid vigor such that when it is grown under the same conditions as comparable high yielding inbred rice varieties it can produce up to 30percent more rice. However, the heterosis effect disappears after the first (F1) generation, so the farmers cannot save seeds produced from a hybrid crop and need to purchase new F1 seeds every season to get the heterosis effect each time.

**Table A2: Baseline Characteristics and Balancing for Hybrid Rice Producing Households**

|  | Means by treatment | | |
| --- | --- | --- | --- |
|  | Control (1) | Treatment (2) | P-value (3) |
| **Variables** | | | |
| **Household Composition** | | | |
| Female headed household | 0.03 | 0.02 | 0.76 |
|  | (0.02) | (0.01) |  |
| Age of household head (in years) | 46.03 | 44.35 | 0.25 |
|  | (1.06) | (1.11) |  |
| Household size (number of members) | 4.72 | 4.69 | 0.93 |
|  | (0.17) | (0.16) |  |
| Number of adult members (>16 years) | 3.33 | 3.07 | 0.12 |
|  | (0.13) | (0.13) |  |
| Number of child (<16 years) | 1.39 | 1.62 | 0.33 |
|  | (0.10) | (0.12) |  |
| Household head with no education | 0.41 | 0.27 | 0.11 |
|  | (0.05) | (0.04) |  |
| **Amount of Land and Credit** | | | |
| Own cultivated land (in decimal[†]) | 50.54 | 55.24 | 0.59 |
|  | (5.10) | (6.51) |  |
| Rented in land (in decimal) | 62.75 | 66.50 | 0.81 |
|  | (7.76) | (10.34) |  |
| Rented out land (in decimal) | 2.56 | 7.49 | 0.10 |
|  | (0.97) | (2.79) |  |
| Total cultivated land (in decimal) | 113.29 | 121.75 | 0.58 |
|  | (8.07) | (11.33) |  |
| Formal and informal loan amount (in taka) | 4669.64 | 3930.69 | 0.80 |
|  | (1363.97) | (1760.58) |  |

*Notes*: [†]A **decimal** (also spelled **decimel**) is a unit of area in India and Bangladesh approximately equal to 1/100 acre (40.46 m²); 247 decimals=1 hectares. [††]**kilogram** (1 kilogram=2.204 pounds). Column 3 shows the P value of mean difference column 3=column1- column2.
**Unit of observation**: Household. Sample includes all rice producing farm households surveyed at baseline (2012). Standard errors (in parentheses) are clustered at Branch level. Informal lenders include moneylenders, loans from friends/family, and buying goods/services on credit from sellers. **Own land** refers to the cultivated crop land owned by the farm household. **Rented in** land means the land rented in from others for crop cultivation. **Rented out** land means the land rented to other farmers for crop cultivation.

# Table A2 (contd.): Baseline Characteristics and Balancing for Hybrid Rice Producing Households

|  | Means by treatment | | |
|---|---|---|---|
|  | Control (1) | Treatment (2) | P-value (3) |
| **Variables** | | | |
| **Output and Input for Rice Production (yearly)** | | | |
| Rice yield (total rice/total land) †† | 18.51 | 16.89 | 0.38 |
|  | (0.57) | (0.48) |  |
| High Yielding Variety rice yield (HYV rice/ land in HYV rice) | 17.66 | 15.48 | 0.18 |
|  | (0.63) | (0.47) |  |
| Hybrid rice yield (HB rice/ land in HB rice) | 20.39 | 19.94 | 0.87 |
|  | (0.73) | (0.65) |  |
| Traditional Variety rice yield (TV rice/ land in TV rice) | 9.60 | 10.21 | 0.54 |
|  | (0.44) | (0.74) |  |
| Total Land (in decimal) | 161.12 | 157.63 | 0.91 |
|  | (12.75) | (11.22) |  |
| Total Labor (days) | 69.14 | 70.32 | 0.93 |
|  | (5.24) | (5.31) |  |
| Total Plough (number of times) | 9.79 | 10.59 | 0.64 |
|  | (0.85) | (0.76) |  |
| Total Seed (kilogram) | 31.66 | 34.62 | 0.50 |
|  | (2.99) | (2.84) |  |
| Total Irrigation (hours) | 55.58 | 66.80 | 0.55 |
|  | (5.02) | (6.26) |  |
| Total Fertilizer (kilogram) | 218.08 | 258.82 | 0.44 |
|  | (18.58) | (19.14) |  |
| Total Pesticide (number of times) | 7.66 | 4.68 | 0.01 |
|  | (0.80) | (0.50) |  |

*Notes*: †† Rice in measured in kilogram (1 kilogram=2.204 pounds). Land is measured in decimal (247 decimals=1 hectares.) Column 3 shows the P value of mean difference column 3=column1- column2.
**Unit of observation**: Household. Sample includes all rice producing farm households surveyed at baseline (2012). Standard errors (in parentheses) are clustered at Branch level. **High Yielding Varieties (HYV)** rice seeds are land substituting, water economizing, more labor using, and employment generating innovation. HYVs significantly outperform traditional varieties in the presence of an efficient management of irrigation, pesticides, and fertilizers. However, in the absence of these inputs, traditional varieties may outperform HYVs.
**Hybrid rice** is any genealogy of rice produced by crossbreeding different kinds of rice. It typically displays heterosis or hybrid vigor such that when it is grown under the same conditions as comparable high yielding inbred rice varieties it can produce up to 30percent more rice. However, the heterosis effect disappears after the first (F1) generation, so the farmers cannot save seeds produced from a hybrid crop and need to purchase new F1 seeds every season to get the heterosis effect each time.

**Table A3: Impact of Access to Credit on Rice Productivity (For HYV and Hybrid Rice)**

| Variables | Effect of Credit Access (1) | Effect of Credit Access (2) | Observations |
|---|---|---|---|
| High Yielding Variety rice yield (HYV rice/ Land in HYV rice) | 12.49*** | 12.61*** | 2,831 |
|  | (0.83) | (0.76) |  |
| Hybrid rice yield (HB rice/ Land in HB rice) | 10.77* | 11.78** | 412 |
|  | (1.37) | (1.09) |  |
| Average rice yield at baseline (Total rice/Total land) | 18.12 | | |
| Control for Baseline covariates | No | Yes | |

*Notes*: ***p<0.01, **p<0.05, *p<0.1. Rice is measured in kilogram (1 kilogram=2.204 pounds). Land is measured in decimal (also spelled **decimel**) which is a unit of area in India and Bangladesh approximately equal to 1/100 acre (40.46 m²); 247 decimals=1 hectares. Column (1) and (2) shows the impact of the treatment on outcome of interest. Standard errors (in parentheses) are clustered at the Branch level.

Table A4: Impact of Credit Access on Frontier Shift and Efficiency of Rice Production (percentage effect)

| Variables | High Yielding Variety Rice Yield (HYV Rice/Land in HYV) | | | | Hybrid Rice Yield (HB Rice/Land in HB) | | | |
|---|---|---|---|---|---|---|---|---|
| | Impact on Frontier Shift | Impact on Frontier Shift | Impact on Inefficiency | Impact on Inefficiency | Impact on Frontier Shift | Impact on Frontier Shift | Impact on Inefficiency | Impact on Inefficiency |
| | (1) | (2) | (3) | (4) | (5) | (6) | (7) | (8) |
| Credit access (1=assigned in treatment group) | 10.89*** | 9.94*** | -1.68 | -3.13** | 2.70 | 3.17 | -9.39** | -8.58** |
| | (1.15) | (1.15) | (1.60) | (1.58) | (3.81) | (3.86) | (3.99) | (3.94) |
| Mean Inefficiency in Rice Production | 13.86 | 13.84 | 13.86 | 13.84 | 17.21 | 16.44 | 17.21 | 16.44 |
| Baseline Covariates | no | yes | no | yes | no | yes | no | yes |
| Observations | 2,187 | 2,187 | 2,036 | 2,036 | 269 | 269 | 269 | 269 |

*Notes*: Unit of observation is household. Sample includes rice producing farm households. Rice in measured in kilogram (1 kilogram=2.204 pounds). Land is measured in decimal which is a unit of area in India and Bangladesh approximately equal to 1/100 acre (40.46 m2); 247 decimals=1 hectares. Standard errors (in parentheses) are clustered at Branch level.

**Table A5: Impact of Access to Credit on Familiarity with Agricultural Extension Officers**

| Variables | Effect of Credit Access | Observations |
|---|---|---|
| Familiarity with Agricultural extension officers[1] (dummy) | 12.15 | 3,172 |
|  | (9.50) |  |
| Discussion about production and farm practices[2] (dummy) | 10.89 | 3,172 |
|  | (9.76) |  |
| Control for Baseline Covariates | Yes |  |

*Notes*: ***p<0.01, **p<0.05, *p<0.1. Standard errors (in parentheses) are clustered at the branch level. [1]Familiarity with agriculture extension service provider implies that farmers are acquainted with the persons/institution from whom they can seek information or advice on crop selection, crop rotation, modern cropping technology, appropriate use of fertilizer, pesticide etc. [2]takes a value of 1 if the farmers discussed or seek advice from the extension service provider on crop selection, crop rotation, modern cropping technology, appropriate use of fertilizer, pesticide etc.

**Table A6: Mean of Demographic and Farm Characteristics**

| Variables | Observations | Mean | Std. Dev. |
|---|---|---|---|
| **Household Composition** |  |  |  |
| Female Headed household (dummy) | 3,172 | 0.05 | 0.21 |
| Head with no education (dummy) | 3,172 | 0.45 | 0.49 |
| Small farm size (1=cultivated land<50 decimal) | 3,172 | 0.27 | 0.44 |
| Pure tenant farm households (1= no own land rice cultivation) | 3,172 | 0.32 | 0.47 |
| Mixed tenant farms (1=cultivate own land as well as others land) | 3,172 | 0.34 | 0.47 |

*Notes*: Unit of observation: Household. Sample includes all rice producing farm households surveyed in 2014.

## Figure B1: Distribution of BCUP Credit

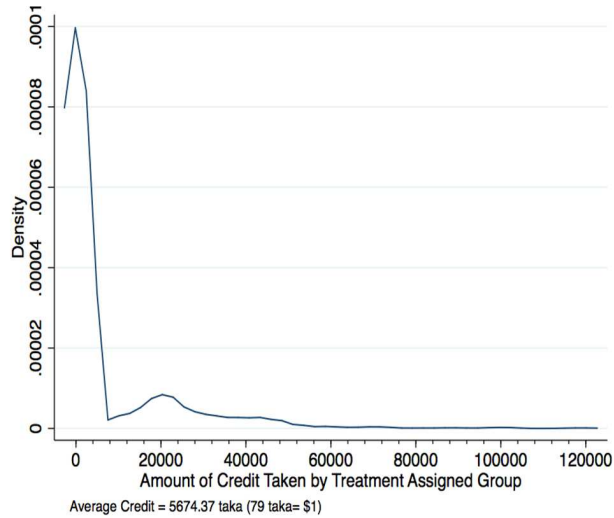

Panel A: Distribution of Credit Amount
(Treatment Assigned Households)

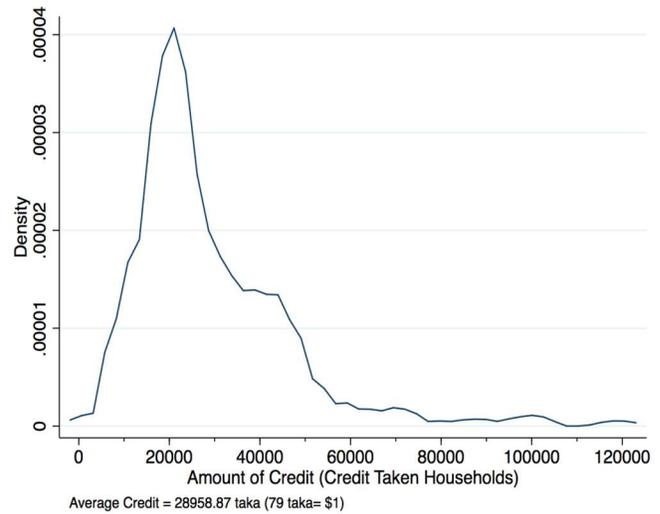

Panel B: Distribution of Credit Amount
(Households who Actually Took Credit)

## Table B1: Impact of Access to Credit Access on Rice Production by Risk Preference

| Variables | Dependent Variable: Rice (in kilogram) | | | |
|---|---|---|---|---|
| | Traditional Variety | Traditional Variety | Hybrid and HYV Variety | Hybrid and HYV Variety |
| | (1) | (2) | (3) | (4) |
| Credit access (1=assigned in treatment group) | -130.81 | -122.87 | 716.74* | 689.38* |
| | (124.6) | (113.13) | (407.45) | (406.96) |
| Risk Averse[1] | 11.94 | 14.43 | -108.54 | -118.07 |
| | (23.25) | (21.06) | (97.64) | (87.54) |
| Risk Averse * Credit Access | 90.84** | 85.23 | -266.13 | -203.72 |
| | (43.55) | (41.31) | (125.69) | (112.87) |
| Baseline Mean Rice Amount | 324.14 | 324.14 | 1614.86 | 1614.86 |
| Baseline Covariates | No | Yes | No | Yes |
| Observations | 2,792 | 2,792 | 2,792 | 2,792 |

*Notes*: Unit of observation is household. Sample includes rice producing farm households. Rice in measured in kilogram (1 kilogram=2.204 pounds). Standard error in parenthesis and are clustered at branch level. *** p<0.01, ** p<0.05, * p<0.1. [1]Risk aversion is gauged by individual's choice over three different type of lottery with same probability of winning and losing. The first one offers low risk and comparatively low return, second lottery offers moderate risk and moderate return, and the third one offers higher benefit and higher risk. Choice of first, second, and the third lottery is assigned a value of 3, 2, and 1 respectively.

**Table B2: Impact of Access to Credit on Borrowing from Different Sources**

| Variables | Dependent Variable: Amount of Borrowing (in taka) | | | | | |
|---|---|---|---|---|---|---|
| | Total | BCUP | Bank and Cooperative | Others NGO | Informal Source | Total Other than BCUP |
| | (1) | (2) | (3) | (4) | (5) | (6) |
| Credit access (1=assigned in treatment group) | 6408.77* | 5639.09*** | 376.16 | -284.2028 | 1336.713 | 769.68 |
| | (3270.21) | (863.86) | (973.01) | (264.0737) | (2415.71) | (3481.96) |
| Baseline Mean | 4914.16 | 0.00 | 1487.56 | 379.71 | 1700.15 | 4914.16 |
| Observations | 4,141 | 4,141 | 4,141 | 4,141 | 4,141 | 4,141 |

*Notes*: Unit of observation is household. Sample includes rice producing farm households. Standard error in parenthesis and are clustered at branch level. *** $p<0.01$, ** $p<0.05$, * $p<0.1$. Informal sources include friends and relatives, traditional moneylenders, landlords etc.

**Table B3: Impact of Access to Credit on the Amount and Type of Land Used for Rice Production**

| Variables | Amon Season | | | Boro Season | | | Total | | |
|---|---|---|---|---|---|---|---|---|---|
| | Amount of land (decimals) | Soil Type (1 if clay or clay-loam) | Land elevation (1 if medium to low land) | Amount of land (decimals) | Soil Type (1 if clay or clay-loam) | Land elevation (1 if medium to low land) | Amount of land (decimals) | Soil Type (1 if clay or clay-loam) | Land elevation (1 if medium to low land) |
| | 1 | 2 | 3 | 4 | 5 | 6 | 7 | 8 | 9 |
| Credit access (1=assigned in treatment group) | -4.23 (8.69) | 0.22* (0.12) | -0.11 (0.08) | -7.24 (8.79) | 0.01 (0.10) | -0.12 (0.07) | -4.31 (15.19) | 0.03 (0.12) | -0.16* (0.09) |
| | | | | | | | | | |
| Baseline Mean | 43.33 | 0.28 | 0.78 | 71.36 | 0.34 | 0.82 | 87.41 | 1700.15 | 4914.16 |
| Observations | 2,769 | 2,779 | 2,779 | 3,278 | 3,282 | 3,282 | 3,857 | 4,107 | 4,107 |

*Notes*: Unit of observation is household. Sample includes rice producing farm households. Standard error in parenthesis and are clustered at branch level. *** p<0.01, ** p<0.05, * p<0.1.